\documentclass[12pt,useAMS]{article}
\pdfoutput=1
\usepackage{natbib209}
\usepackage{psfig}
\usepackage{epsfig}
\usepackage{graphicx}
\usepackage{references}
\newcommand{\be}{\begin{equation}}
\newcommand{\ee}{\end{equation}}
\newcommand{\ba}{\begin{eqnarray}}
\newcommand{\ea}{\end{eqnarray}}
\newcommand{\brr}{\begin{array}}
\newcommand{\err}{\end{array}}
\newcommand{\bc}{\begin{center}}
\newcommand{\ec}{\end{center}}
\newcommand{\hm}{\,h^{-1}{\rm Mpc}}

\newcommand{\msun}{\,h^{-1}M_\odot}

\newcommand{\lum}{\,{\rm erg\,s^{-1}}}

\newcommand{\mincir}{\raise
  -2.truept\hbox{\rlap{\hbox{$\sim$}}\raise5.truept \hbox{$<$}\ }}
\newcommand{\magcir}{\raise
  -2.truept\hbox{\rlap{\hbox{$\sim$}}\raise5.truept \hbox{$>$}\ }}
\newcommand{\siml}{\raise
  -2.truept\hbox{\rlap{\hbox{$\sim$}}\raise5.truept \hbox{$<$}\ }}
\newcommand{\simg}{\raise
  -2.truept\hbox{\rlap{\hbox{$\sim$}}\raise5.truept \hbox{$>$}\ }}

\begin{document}

\title{{\Large \bf Cosmological simulations of galaxy clusters}}

\date{    }
\maketitle
\begin{center}
\vskip -1.5truecm

{Stefano Borgani$^{1,2,3}$ \& Andrey Kravtsov$^{4,5,6}$}\\
\vskip 0.3truecm

{\small 
1- Dipartimento di Astronomia dell'Universit\`a di Trieste, via
  Tiepolo 11, I-34131 Trieste, Italy (borgani@oats.inaf.it)\\
2- INAF, Osservatorio Astronomico di Trieste, via Tiepolo 11,
  I-34131 Trieste, Italy\\
3- INFN -- National Institute for Nuclear Physics, Trieste,
  Italy\\ 
4- Kavli Institute for Cosmological Physics, The University of
Chicago, Chicago, IL 60637, USA (andrey@oddjob.uchicago.edu)\\
5- Enrico Fermi Institute, The University of Chicago, Chicago, IL
60637, USA\\
6- Dept. of Astronomy \& Astrophysics, The University of Chicago, 5640
S. Ellis Ave., Chicago, IL 60605, USA
}
\end{center}

\begin{abstract} 
  We review recent progress in the description of the formation and
  evolution of galaxy clusters in a cosmological context by using
  state-of-art numerical simulations. We focus our presentation on the
  comparison between simulated and observed X--ray properties, while
  we will also discuss numerical predictions on properties of the
  galaxy population in clusters, as observed in the optical band.
  Many of the salient observed properties of clusters, such as scaling
  relations between X--ray observables and total mass, radial profiles
  of entropy and density of the intracluster gas, and radial
  distribution of galaxies are reproduced quite well. In particular,
  the outer regions of cluster at radii beyond about 10 per cent of
  the virial radius are quite regular and exhibit scaling with mass
  remarkably close to that expected in the simplest case in which only
  the action of gravity determines the evolution of the intra-cluster
  gas. However, simulations generally fail at reproducing the observed
  ``cool core'' structure of clusters: simulated clusters generally
  exhibit a significant excess of gas cooling in their central
  regions, which causes both an overestimate of the star formation in
  the cluster centers and incorrect temperature and entropy
  profiles. The total baryon fraction in clusters is below the mean
  universal value, by an amount which depends on the cluster-centric
  distance and the physics included in the simulations, with
  interesting tensions between observed stellar and gas fractions in
  clusters and predictions of simulations. Besides their important
  implications for the cosmological application of clusters, these
  puzzles also point towards the important role played by additional
  physical processes, beyond those already included in the
  simulations. We review the role played by these processes, along
  with the difficulty for their implementation, and discuss the
  outlook for the future progress in numerical modeling of clusters.
\end{abstract}

~~~~~~~~~~~~~~~~keywords: astrophysics, cosmology, computer science,
fluid dynamics

\section{Introduction}
\subsection{Galaxy clusters in the hierarchy of cosmic structures}

Astronomical observations over the last century have revealed the
existence of a continuous hierarchy of cosmic structures involving a
wide range of scales. On the scales of hundreds of thousands of
parsecs\footnote{A parsec (pc) is $3.086\times 10^{18}$~cm.} the
distribution of luminous matter is highly non-uniform and is
concentrated in the ``islands'' which we call galaxies. Galaxies
themselves are not distributed randomly but are concentrated in groups
of two or more galaxies (roughly a million parsec in size), large
concentrations of hundreds and sometimes thousands of galaxies (up to
three--four million pc in size), which we call galaxy clusters, and a
network of filaments (tens of millions of pc in size) interconnecting
individual galaxies, groups, and clusters.

Clusters of galaxies occupy a special position in this hierarchy: they
are the largest objects that have had time to undergo gravitational
collapse. Existence of clusters of nebulae have been known for more
than a hundred years. However, only after Edwin Hubble's proof that
spiral and elliptical nebulae were bona fide galaxies like the Milky
Way located at large distances from us \citep{hubble25,hubble26}, it
was realized that clusters of nebulae are systems of enormous size and
mass. Back in the 1930s astronomers recognized that some invisible
dark matter (DM) should dominate the overall gravitational field of
clusters in order to explain unexpectedly high velocities of galaxies
within the Virgo Cluster, thus making galaxy clusters ``ante
litteram'' cosmological probes (\citealt{zwicky33,smith36,zwicky37};
see \citealt{2000cucg.confE...1B} for a historical overview). The
interest in galaxy clusters has rapidly increased thereafter and large
catalogs of clusters have been constructed using visual searches of
deep photographic sky images for strong concentrations of galaxies
\citep{abell58,zwicky61}. Clusters remain powerful cosmological probes
and astrophysical laboratories today, as larger and more sensitive
surveys of clusters are constructed.

\subsection{Galaxy clusters as cosmological probes and astrophysical
  laboratories} 

Measurements of the velocities of galaxies in clusters have provided
the first determination of the typical mass involved in such
structures, which typically falls in the
range\footnote{$M_{\odot}\approx 1.99\times 10^{33}$~g is the mass of
  the Sun.}  $10^{14}$--$10^{15}M_\odot$. Stars and gas in galaxies,
however, constitute only a small fraction ($\approx 2$ per cent) of
this mass \citep{lin_etal03,gonzalez_etal07}. With the advent of
X--ray astronomy in the late 60s, galaxy clusters have been also
identified as powerful emitters of photons having energies of several
kilo-electronvolts (keV;
\citealt{1971ApJ...167L..81G,1971ApJ...165L..49K}), with typical
luminosities of about $10^{43}$--$10^{45}\lum$. This emission was
interpreted as due to the presence of a
hot, fully-ionized thermal plasma with temperature of several keV (1 keV$\simeq 1.16\times 10^{7}$K) and a typical
particle number density of $10^{-1}$--$10^{-4}$cm$^{-3}$, with the two main contributions to emission from the 
free-free interactions of electrons and ions and line
emission by ions of heavy elements such as iron.
This diffuse plasma is not associated with individual galaxies and
constitutes the intra-cluster medium (ICM), which contains the bulk of
the normal baryonic matter in clusters. The temperature of the ICM is
consistent with velocities of galaxies and indicates that both
galaxies and gas are in equilibrium within a common gravitational
potential well.  The mass of galaxies and hot gas is not sufficient to
explain the implied depth of the potential well, which implies that
most of the mass in clusters is in a form of dark matter.  Given that
hydrogen is by far the most abundant element in the universe, most of
the plasma particles are electrons and protons, with a smaller number
of helium nuclei.  There are also trace amounts of heavier nuclei some
of which are only partially ionized; the typical abundance of the
heavier elements is about a third of that found in the Sun or a
fraction of one per cent by mass.

Since its discovery, the X--ray emission serves as a highly efficient
and clean way to identify galaxy clusters out to large cosmological
distances \citep{2002ARA&A..40..539R,2008arXiv0805.2207V}. In
addition, the population of thermal electrons also leaves its imprint
on the Cosmic Microwave Background (CMB), when observed in the
direction of galaxy clusters: their inverse Compton scattering off the
CMB photons induces small but measurable distortions in the
black--body spectrum of the CMB, equivalent to temperature variations
of about $10^{-4}$--$10^{-5}$ K, thus giving rise to the 
Sunyaev--Zeldovich effect \citep[SZE;][for a 
review see
\citeauthor{carlstrom_etal02}
\citeyear{carlstrom_etal02}]{sz70,sz72,sz80}. Given that the SZE
measurement is simply a measurement of the intensity or temperature
fluctuation of the CMB, it is nearly distance independent, which makes
it a powerful way to detect distant clusters. Indeed, a number of
observational campaigns are now started or planned to search and study
galaxy clusters using the SZE within a significant fraction of the
cosmological past light-cone
\citep{kosowsky03,ruhl_etal04,2008arXiv0810.1578S}.

\begin{figure}
 \hspace{-0.6cm} \centerline{ 
\psfig{file=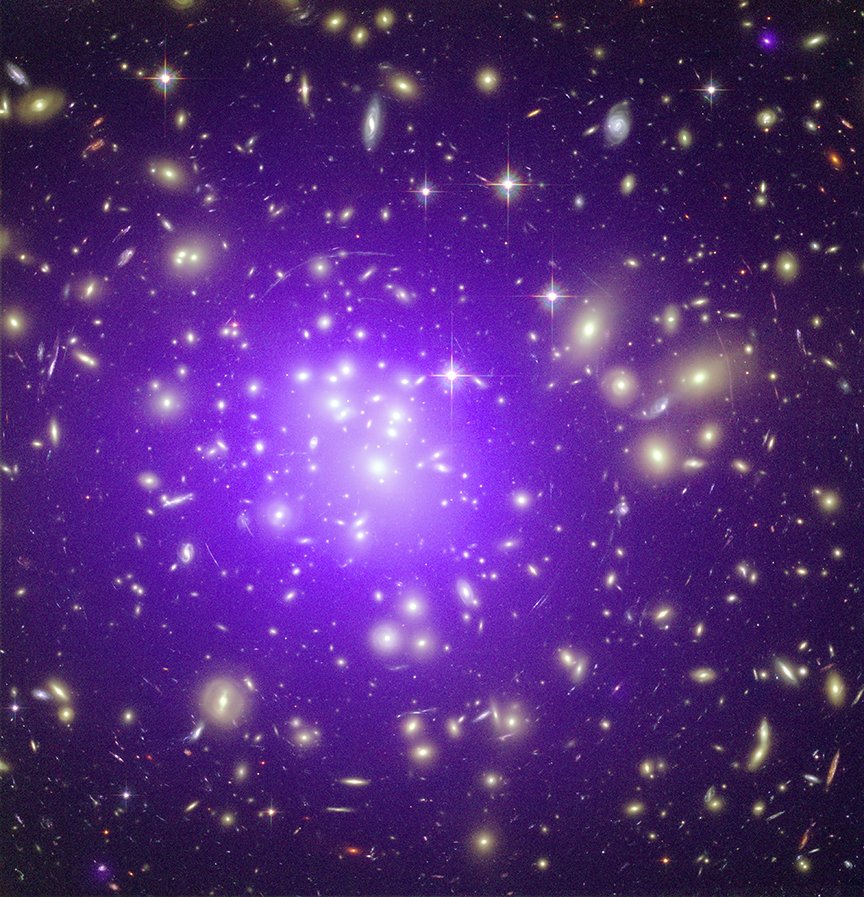,width=8.truecm}
\psfig{file=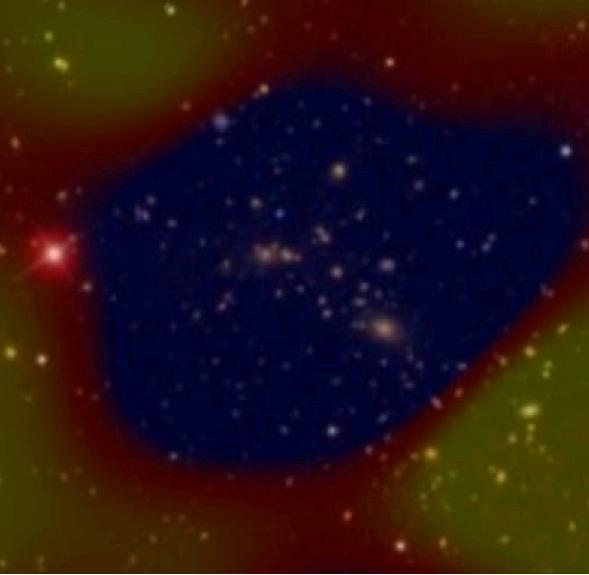,width=8.5truecm} 
}
  \caption{Left panel: composite X--ray/optical image of the galaxy
    cluster Abell 1689, located at redshift $z=0.18$. The map shows
    an area of 556 kpc on a side. The purple diffuse halo shows the
    distribution of gas at a temperature of about $10^8$ K, as
    revealed by the Chandra X-ray Observatory. Images of galaxies in
    the optical band, colored in yellow, are from observations
    performed with the Hubble Space Telescope. The long arcs in the
    optical image are caused by gravitational lensing of background
    galaxies by matter in the galaxy cluster, the largest system of
    such arcs ever found (Credit:X-ray: NASA/CXC/MIT; Optical:
    NASA/STScI). Right panel: optical image of cluster Abell 1914 from
    the Sloan Digital Sky Survey with the superimposed map of the
    temperatures of the Cosmic Microwave Background observed by the
    Sunyaev-Zeldovich array (SZA). The image illustrates the effect
    of up-scattering of the CMB photons by the hot ICM from low
    frequencies to higher frequencies. At the frequency of
    observation, the cluster appears as a temperature decrement in the
    CMB temperature map. (Credit: John Carlstrom and SZA collaboration).
  }
\label{fig:a1689}
\end{figure}

Figure \ref{fig:a1689} shows an overlay of the optical, X--ray and SZE
images of the massive clusters Abell 1689 and 1914.
It illustrates all of the main components of the clusters: the luminous stars, the hot
ICM observed via its X-ray emission and the SZE, and even the presence
of the unseen dark matter manifesting itself through gravitational
lensing of background galaxies (see below). The figure shows several
bright elliptical galaxies that are typically located near the cluster
centers. A salient feature of such central galaxies is that they show
little evidence of ongoing star formation, despite their extremely
large masses. Furthermore, a large amount of DM, extending well beyond
the region traced by the X--ray emission, leaves its imprint in the
pattern of gravitational lensing, which causes the distortion of the
images of background galaxies. In the inner regions of clusters the
 gravitational lensing is strong and its effects can be easily seen in the distorted 
images of background galaxies appearing as long thin arcs curved
around the cluster center. At larger radii, the effect is
weaker. Although not easily visible by eye, it can still be reliably
measured by estimating shapes of many background galaxies and
comparing their statistical average with the expected value for an
isotropic distribution of shapes. The gravitational lensing is direct
probe of the total mass distribution in clusters, which makes it both
extremely powerful in its own right and a very useful check for other
methods for measuring cluster masses.

Given their large mass, galaxy clusters represent the end result of
the collapse of density fluctuations involving comoving scales of
$\sim 10$ Megaparsecs (Mpc). For this reason, they mark the transition
between two distinct dynamical regimes. On scales roughly above 10
Mpc, the evolution of the structure of the universe is mainly driven
by gravity. In this regime, the evolution still feels the imprint of
the cosmological initial conditions and can be described by analytical
methods and, more accurately, by cosmological N-body simulations. The
latter follow the growth of seed fluctuations in the matter density
field from an early epoch to the present time (or even beyond) due to
gravitational instabilities.

\begin{figure}
\centerline{
\psfig{file=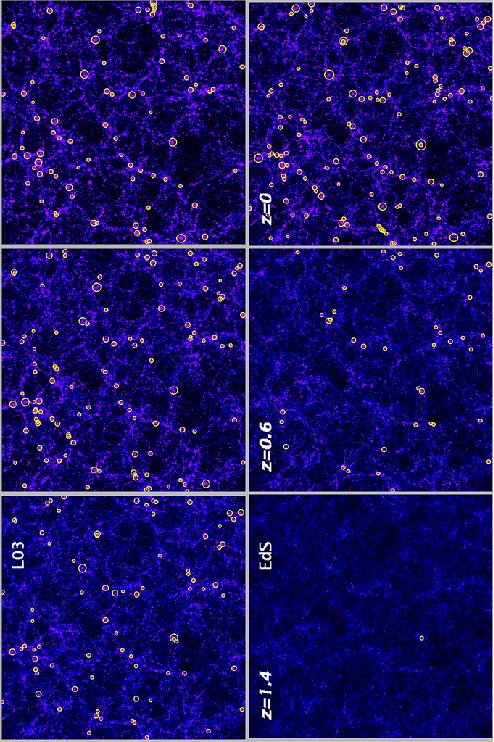,angle=270,width=17.truecm}
}
\caption{Evolution of large-scale cosmic structures simulated in
N-body simulations of two different cosmological models. Each of the three
redshift snapshots shows a region with 250$\hm$ side and 75$\hm$ thick
(co-moving scale-lenghts).  The upper panels describe a flat
low--density model with $\Omega_{\rm m}=0.3$ and cosmological constant
contribution to the energy density of $\Omega_\Lambda=0.7$, while the
lower panels are for another spatially flat cosmological model with
$\Omega_{\rm m}=1$. In both cases the amplitude of the power spectrum
is consistent with the number density of nearby galaxy
clusters. Superimposed on the matter distribution, the yellow circles
mark the positions of galaxy clusters that would be seen shining in
$X$--rays with a temperature $T>3$ keV. The size of the circles is
proportional to temperature.  The difference in the evolution of
cluster abundance in the two models illustrates the importance of
clusters as probes of the dark matter and dark energy content of the
universe. Figure adopted from \protect\cite{2001Natur.409...39B} with
copyright permission from 2001 Nature Publishing Group.}
\label{fig:clusevol}
\end{figure}

An example of how galaxy clusters trace cosmic evolution in
cosmological simulations can be appreciated in Figure
\ref{fig:clusevol}, which shows the evolution of the population of
galaxy clusters, superimposed on the evolution of the matter density
field for two different cosmological models. The upper panels are for
a ``concordance'' spatially flat $\Lambda$CDM model, in which the mean
mass density, in units of the critical density of the
universe\footnote{The critical density, $\rho_c(z)=3H(z)^2/8\pi
  G\approx 2.77\times 10^{11} (h^{-1}M_\odot)/(h^{-1}{\rm
    Mpc})^3\approx 1.88\times 10^{-29}\,h^2\rm\, g\,cm^{-3}$,
  corresponds to the density for which the Universe is spatially flat
  and separates the models which will expand forever from those that
  will re-collapse at some point in the future. Here and in the
  following of this paper, $h$ denotes the value of the Hubble
  constant at $z=0$ in units of 100 $\rm km\, s^{-1}Mpc^{-1}$}, is
$\Omega_{\rm m}=0.3$ and cosmological constant contributes the rest of
the energy density required to make the universe flat,
$\Omega_{\Lambda}=0.7$. The lower panels also show a flat model, but
without any cosmological constant (i.e., matter density is
$\Omega_{\rm m}=1$). Parameters of the two models have been chosen to
produce a comparable number of clusters at the present time. Quite
apparently, the number density of clusters evolves much more slowly in
the $\Lambda$CDM model. Indeed, while this model predicts a
significant population of clusters at redshift $z>1$, and minimal
evolution between $z=0.6$ and $z=0$, the abundance of such distant
clusters rapidly and steadily drops in the $\Omega_m=1$ Einstein-de
Sitter model at $z>0$.  This is mainly because universe expands faster
in models with lower mean matter density $\Omega_{\rm m}$. 
For geometrically flat cosmological models with low mean
matter density, $\Omega_m<1$, the cosmological constant starts to
dominate the energy density of the universe and drive accelerating
expansion at redshift $(1+z)=\Omega_m^{-1/3}$ or $z\approx 0.5$
\citep[e.g.,][]{carroll_etal92}.  This example highlights the
important role that galaxy clusters play in tracing the cosmic
evolution and in constraining the dark matter and dark energy content
of the universe.

At scales below 1 Mpc, the physics of baryons starts
to play an important role in addition to gravity, thus significantly
complicating the associated processes.  As we describe in more detail
below, in the current paradigm of structure formation clusters are
thought to form via a hierarchical sequence of mergers and accretion
of smaller systems driven by gravity and dark matter that dominates
the gravitational field. During this sequence the intergalactic gas is
heated to high, X-ray emitting temperatures by adiabatic compression
and shocks, and settles in hydrostatic equilibrium within the cluster
potential well. Once the gas is dense enough, it cools, leaves the hot
phase, forms the stellar component and can accrete onto supermassive
black holes (SMBHs) harbored by the massive cluster galaxies. The
process of cooling and formation of stars and SMBHs can then result in
energetic feedback due to supernovae or 
active galactic nuclei (AGN),
which can inject substantial amounts of heat into the ICM and spread
heavy elements throughout the cluster volume.

Within this global picture, galaxy clusters represent the place where
astrophysics and cosmology meet each other: while their overall
dynamics is dominated by gravity, the astrophysical processes taking
place at the galactic scale leave observable imprints on the diffuse
hot gas \citep[e.g.,][]{voit05}.  Indeed, the gravitational potential
wells of massive clusters are extremely deep and clusters therefore
are expected to contain a universal fraction of baryons within a
suitably large radius\footnote{As we discuss in \S~\ref{sec:gasfrac}
  below the observational picture is currently more complicated.}.
Given that we can probe each of their components with observations,
clusters are powerful laboratories where to study the processes
operating during galaxy formation and their effects on the surrounding
intergalactic medium.  As we will discuss in the following sections,
hydrodynamical codes coupled to N-body techniques represents the most
advanced instruments to describe such complex processes and their
impact on the assembly history of cosmic structures.

\subsection{Theoretical models of galaxy clusters}

\subsubsection{The ``spherical cow'' model: self--similar clusters}
\label{sec:selfsimilar}
Before we delve into the complexities of the numerical description of
cluster formation, it is instructive to consider the simplest model,
which provides a useful baseline for evaluation of more accurate and
sophisticated models.  Such a model assumes that cluster properties
and correlations between them are determined by gravity alone and that
clusters are in virial equilibrium \citep[][see \citeauthor{voit05}
\citeyear{voit05} for a review]{kaiser86}. Since gravity does
not have preferred scales, in this model we expect clusters to be {\it
  self-similar\/}
(i.e., clusters of different mass to be the scaled version of each
other) and their mass to be the only parameter that determines the
thermodynamical properties of the intra-cluster gas.

If, at redshift $z$, we define $M_{\Delta_c}$ to be the mass contained
within the radius\footnote{Throughout this paper the subscript of the
  radius indicates either the overdensity which the radius encloses
  with respect to the critical density $\rho_c(z)$ at the redshift of
  observation (e.g. $r_{500}$ is the radius enclosing overdensity of
  $500\rho_c$) or that the radius encloses the virial overdensity
  ($r_{\rm vir}$) predicted by the spherical collapse model
  \citep[e.g.][]{eke_etal96}.} $r_{\Delta_c}$, encompassing a mean
overdensity $\Delta_c$ times the critical cosmic density at the
redshift of observation $\rho_c(z)$, then $M_{\Delta_c} \propto
\rho_c(z) \Delta_c r_{\Delta_c}^3$.  The critical density of the
universe scales with redshift as $\rho_c(z)=\rho_{c0} E^2(z)$, where
\be
E(z)\,=\,H(z)/H_0\,=\,\left[(1+z)^3\Omega_m+(1+z)^2\Omega_k+\Omega_\Lambda\right]^{1/2} 
\ee
gives the evolution of the Hubble parameter $H(z)$. In the above
expression $\Omega_k=1-\Omega_m-\Omega_\Lambda$ is
the contribution from curvature \citep[we neglect here any contribution from
relativistic species, e.g.,][]{peebles93}. Therefore, the cluster
size $r_{\Delta_c}$ scales with $z$ and $M_{\Delta_c}$ as
$r_{\Delta_c}\propto M_{\Delta_c}^{1/3} E^{-2/3}(z)$.

Given that the gas heated by gravitational infall is assumed to be in 
equilibrium within the gravitational potential $\Phi$ of the cluster,
it is expected to have the temperature $k_BT\propto \Phi\propto
M_{\Delta_c}/r_{\Delta_c}$ ($k_B$: Boltzmann constant). Therefore the
above relation between radius and mass gives
\begin{equation}
M_{\Delta_c}\,\propto \,T^{3/2}E^{-1}(z)\,.
\label{eq:mt_ss}
\end{equation}
This simple relation between mass and temperature can be turned into
scaling relations among other observable quantities.  

As for the X-ray luminosity, it scales as the characteristic emissivity
times the volume occupied by the cluster. If we assume that thermal
bremsstrahlung process dominates the emission from the ICM plasma with
electron number density $n_e$
\citep[e.g.,][]{sarazin86,peterson_fabian06}, we have $L_X\propto
n_e^{2}T^{1/2}r^3_{\Delta_c}$. The characteristic density scales as
$n_e\propto M/r^3_{\Delta_c}={\rm const}$ and $T\propto
M_{\Delta_c}/r_{\Delta_c}$, which (using the $M-T$ relation above)
gives $L_X\propto T^2E(z)$.

Another useful quantity to characterize the thermodynamical properties
of the ICM is the entropy \citep{voit05}. In X--ray studies of the
ICM, it is usually defined as
\be
S\,=\,{k_BT\over \mu m_p \rho_{gas}^{2/3}}\,.
\label{eq:entrK} 
\ee
With the above definition, the quantity $S$ is the constant of
proportionality in the equation of state of an adiabatic mono-atomic
gas, $P=S\rho_{gas}^{5/3}$. Using the thermodynamic definition of
specific entropy, $s=c_V\ln(P/\rho_{gas}^{5/3})$ ($c_V$: heat capacity
at constant volume), one obtains $s=k_B\ln S^{3/2}+s_0$, where $s_0$
is a constant. Another quantity, often called ``entropy'' in cluster studies, that we
will also use in the following, is
\be
K\,=\,k_B T n_e^{-2/3}\,.
\label{eq:entr}
\ee 
According to the self--similar model, this quantity, computed at a
fixed overdensity $\Delta_c$, scales with temperature and redshift
according to
\be
K_{\Delta_c}\propto T E^{-4/3}(z)\,.
\label{eq:entr_ss}
\ee

Self--similar X--ray scaling relations for the ICM are also predicted
by the spherically-symmetric accretion model originally proposed by
\cite{1985ApJS...58...39B}. Supersonic
accretion gives rise to an expanding accretion shock taking place at
the interface of the inner hydrostatic gas with a cooler,
adiabatically compressed, external medium. This model has been later
generalized to include cooling of gas (with cooling functions of a certain form)
\citep{abadi_etal00} and  to include
the effect of external pre--heating of the diffuse inter-galactic
medium, still under the assumption of spherical accretion, as a way to break
self--similarity \citep{2001ApJ...546...63T,2003ApJ...593..272V}.

The predictions of the self--similar model have been tested by a
number of authors against hydrodynamical simulations, which include
only the effect of gravitational heating
\citep[e.g.,][]{navarro_etal95,eke_etal98,nagai_etal07b}. These
simulations generally confirmed the above scaling relations for the
overall average quantities and for the radial profiles, as well as
their evolution with redshift, although small deviations from
self--similarity, due to the differences in formation redshifts of
systems of different mass and some small differences in the dynamics
of baryons and dark matter have also been found
\citep{ascasibar_etal06}. As we shall discuss in
\S~\ref{sec:morereal}, however, the predictions of this model are at
variance with respect to a number of X--ray observations, 
indicating importance of other physical processes in addition
to gravitational heating.

\subsubsection{Numerical simulations of cluster formation}

The self--similar model and its more complicated
spherically-symmetric extensions introduced in the previous section,
while useful as a baseline for the interpretation of observations, are far
too simple to capture all of the complexities of cluster formation
(see \S~\ref{sec:nonradprop} and \ref{sec:morereal}). As already
mentioned, numerical
cosmological simulations carried out on massive parallel
supercomputers represent the modern instruments to describe these
complexities. Such simulations start at a sufficiently early epoch
when density fluctuations are still small and can be specified using
early evolution models. The initial conditions are a realization of a
density field with statistical properties\footnote{The initial density
  field is usually assumed to be Gaussian \citep[see, however,][for
  examples of non-Gaussian
  simulations]{grossi_etal08,dalal_etal08,2008arXiv0811.4176P}, 
  in which case it is fully specified by the power spectrum of
  fluctuations.} (e.g., the power spectrum) appropriate for the
adopted background cosmological model. The numerical simulations
follow the co-evolution of the collisionless dark matter component and
normal baryonic matter, which interact only via gravity, starting from
the initial conditions and advancing the density and velocity fields
forward by numerically integrating equations governing the dynamics of
dark matter and baryons. 

A number of observational probes, such as the Cosmic Microwave
Background, the statistics of the population of galaxies and galaxy
clusters, the Hubble diagram of Type Ia supernovae, 
the pattern of gravitational lensing, and the
properties of absorption systems in the spectra of distant quasars,
have provided tight constraints on the underlying cosmological model
\citep[e.g., ][ and references therein]{2008arXiv0803.0547K}. This
implies that initial conditions for cosmological simulations can now
be fixed with a far lower degree of ambiguity than it was possible
until ten years ago. Cosmological simulations are therefore quite
"lucky" compared to many other areas of computational science in that
the initial conditions are unique and well specified.  The main
challenge for the simulations is thus to faithfully follow dynamics of
matter driven by gravitational instability and the gas-dynamical
processes affecting the evolution of the cosmic baryons.

The dynamics of collisionless DM is described by the collisionless
Boltzmann equation (also known as the Vlasov equation), which is the
continuity equation of the fine-grained phase space density in the six
dimensional space of coordinates and velocities. The
high dimensionality of the phase space makes it extremely demanding to
numerically solve the Boltzmann equation directly. Therefore, as is
often the case in integrating equations of high dimensionality, the
Monte Carlo technique is used. The solution to the Vlasov equation can
be represented in terms of characteristic equations, which describe
lines in phase space along which the distribution function is
constant.  These equations are identical to the standard Newtonian
equations of particle motion. The complete set of characteristic
equations is equivalent to the Vlasov equation. In cosmological
simulations only a representative subset of characteristic equations
is solved by discretizing and sampling the initial phase space by $N$
particles ("bodies") and then integrating their equations of motion in
the collective gravity field (equivalent to solving the characteristic
equations).

Dynamics of diffuse baryonic matter is expected to be highly collisional 
(see \S~\ref{sec:addphysics}) and is therefore followed using standard 
hydrodynamics techniques. 
All of the matter components in a simulation interact with each
other only via gravity.  The integration of a coupled set of the
governing equations allows one to provide information about matter
properties in three dimensional space at different epochs.

Crucial parameters in defining the accuracy of a simulation are the
mass and spatial resolution.  The mass resolution is intuitive: it is
simply the mass of the smallest mass element. Spatial resolution
requires some explanation. Each particle or volume element in a
simulation is assumed to have a certain shape (for example, a box or
a sphere), size and internal density distribution. For the
non-overlapping cells, used for example in the Eulerian hydrodynamics
solvers, the spatial resolution is simply the size of the smallest
cells. For collisionless dark matter particles or particle-based
hydrodynamics solvers, in which particles can overlap, the resolution
is the scale at which gravity and/or hydrodynamics forces are
smoothed.

The mass and spatial resolution should be tightly related
\citep[e.g.,][]{knebe_etal00,power_etal03,diemand_etal04}, but are not
always identical in the actual simulations (e.g., high mass resolution
in a simulation with low spatial resolution does not necessarily mean
that the smallest mass elements are faithfully followed). Ideally,
increasing the number of particles and fluid elements employed to
describe the dynamics should lead to convergence to the true
solution. In practice, however, the convergence can be slow and the
convergent solution may deviate from the true solution because
discretized versions of equations solved in simulations are different
from the real equations (e.g., due to the addition of extra terms
related to numerical viscosity, etc.).

The first simulations of cluster formation have been performed in the
1970s and 1980s \citep{peebles70,white76} and were instrumental in
testing the general idea of gravitational instability as the driving
process of cluster formation. The first simulations of galaxy
clusters, which followed the dynamics of both baryons and dark matter,
were carried out during the late 80s and early 90s
\citep{evrard88,evrard90,thomas_couchman92,katz_white93,bryan_etal94,kang_etal94,navarro_etal95}. Many
of these pioneering simulations have assumed that radiative losses of
the diffuse gas can be neglected, since cooling time for the
bulk of the gas is longer than a typical cluster age or even the age
of the universe.

The addition of baryonic component has allowed simulations to predict
properties of clusters observable in X-rays, as they confirmed that
gas was heated to X-ray emitting temperatures during cluster formation
via adiabatic compression and shocks. These early simulations were
quite successful in reproducing general morphological characteristics
of the X-ray observations available at the time. The simulations have
also showed that gas in the inner regions of clusters is in
approximate hydrostatic equilibrium and have confirmed the expectation
from the simple models that their X-ray observables are tightly
correlated with the total cluster mass \citep{evrard_etal96}.  These
advances provided a foundation for the use of clusters as probes of
cosmological parameters. Further progress in simulations and
observations has shown that details of cluster formation and physics
are more complicated and involve effects related to formation end
evolution of cluster galaxies. While the outer regions of clusters, at
radii larger than about 10 per cent of the virial radius, are quite
regular, nearly self-similar, and are well described by simulations,
the inner core regions exhibit large scatter and evidence of galaxy
formation effects and energy injection by the central AGN. Studies of these
effects have blossomed into an active and vibrant field of research. We
review the numerical modeling aspect of this field in the following
sections.

\section{Techniques of cosmological simulations of cluster formation}
\label{sec:techn}

As we mentioned in the previous section, cosmological simulations of
cluster formation are started from the initial density field at an
early epoch, when the typical matter density fluctuations
around the cosmic mean density are small ($\delta\rho/\rho<0.3$) at
the smallest resolved scales\footnote{The scale is
  comparable to the mean interparticle separation}. The fiducial
assumption is that the density field is Gaussian with a power spectrum
of fluctuations that can be exactly computed for the assumed
cosmological model \citep[see, e.g.,][]{seljak_etal03}. The field is
evolved to the chosen initial epoch using a model sufficiently
accurate to describe the evolution of fluctuations in linear and
mildly non-linear regime, such as the Zeldovich approximation
\citep[][for recent reviews see \citeauthor{sirko05}
\citeyear{sirko05} and \citeauthor{prunet_etal08}
\citeyear{prunet_etal08}]{zeldovich70,klypin_shandarin83,efstathiou_etal85}
or higher-order perturbation theory \citep[e.g.,][]{crocce_etal06}.

In cosmological simulations, gravitational forces acting on both dark
matter and baryons and hydrodynamics forces acting only on baryonic
matter are computed using different numerical schemes. For gravity,
the most direct scheme is based on computing the force among each pair
of particles. Although this direct integration scheme
\citep[e.g.][]{2001NewA....6..277A} is in principle the most accurate,
the required number of operations scales like the square of the total
number of employed particles, thus making it prohibitively expensive
for large cosmological simulations. For this reason, different schemes
have been developed to trade between computational speed and numerical
resolution or accuracy.
Such schemes include grid-based particle-mesh (PM) and
particle-particle/particle-mesh (P$^3$M) methods
\citep{hockney_eastwood88, couchman91}, the gridless tree method where
forces are computed by multipole expansion \citep{barnes_hut86,
  bouchet_hernquist88}, or a hybrid of the two methods with the fast
PM scheme used to compute gravitational forces on large scales and the
tree algorithm to compute forces on smaller scales
\citep{2002JApA...23..185B,bode_etal03,springel_etal05}. The PM
methods solve the Poisson equations discretized either on uniform or
non-uniform grids (e.g. in the Adaptive Mesh Refinement methods) using
FFT method or a multi-grid relaxation solver. Each particle is
treated as a cubic volume element with a specified inner density
distribution when its density is interpolated onto the grid for the
Poisson solver. Gravitational forces are usually
computed by finite differencing the potential
produced by the Poisson solver and are used to advance particle
positions and velocities or update baryon variables in a grid cell.

As for hydrodynamics\footnote{The validity of hydrodynamic treatment
  of intracluster plasma as an ideal fluid is a rather complicated
  issue, which we discuss in \S~\ref{sec:addphysics}.}, the two most
commonly used approaches are gridless smoothed particle hydrodynamics
(SPH) and shock-capturing grid-based methods\footnote{Other schemes,
  such as Godunov-type Particle Hydrodynamics
  \citep[GPH;][]{2002JCoPh.179..238I}, Smoothed Lagrangian
  Hydrodynamics \citep[SLH;][]{gnedin95} and Adaptive Moving Mesh
  \citep{pen98}.}. In the SPH, fluid elements describing the system
are sampled and represented by particles, and the dynamic equations
are obtained from the Lagrangian form of the hydrodynamic conservation
laws \citep[see ][for a recent review]{2005RPPh...68.1703M}. The main
advantage of SPH lies in its Lagrangian nature as there is no grid to
constrain the dynamic range in spatial resolution or the global
geometry of the modeled systems.  However, in order to prevent
particles from penetrating shock regions, the SPH scheme uses an
artificial viscosity, which allows the kinetic energy of such
particles to be converted into thermal energy. Clearly, the presence
of an artificial viscosity term could prevent the development of shear
motions, thus spuriously preventing the development of turbulent
flows. This limitation of the SPH codes can be at least partially
overcome by introducing suitable schemes to let artificial viscosity
decay away from the shock regions
\citep{morris_monaghan97,dolag_etal05}.  Furthermore, SPH has a
limited ability to describe strong gradients in thermodynamical
variables (occurring, for example, in a multi-phase medium or during
cluster mergers; \citealt{agertz_etal07}) or low-density regions with
a finite number of particles.

\begin{figure}[t]
\vspace{-0.5cm}
\centerline{ 
   \hspace{2cm}\epsfysize=4truein  \epsffile{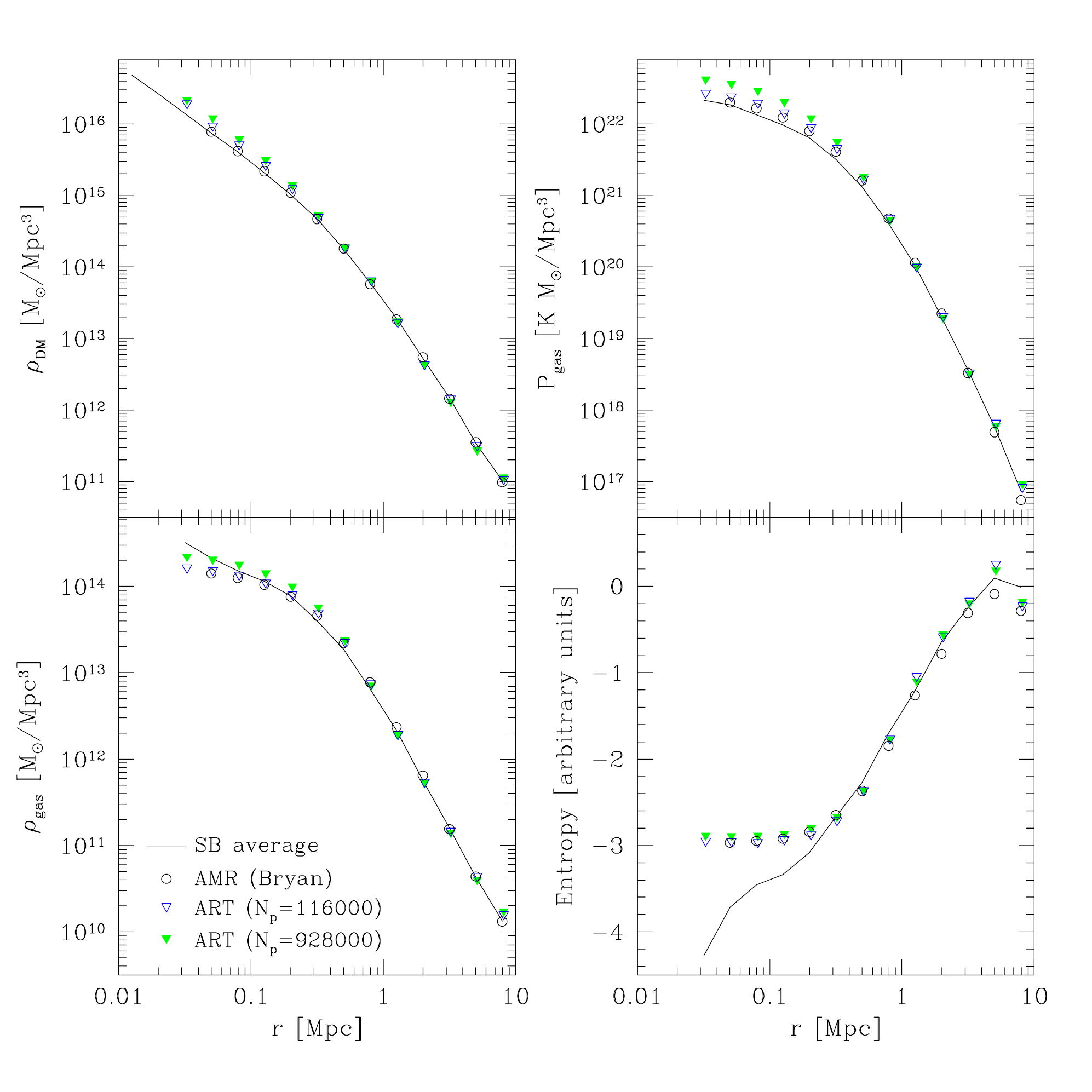}
}
\vspace{-0.5cm}                                                     
\caption{
  A comparison between profiles of dark matter density, gas density,
  gas pressure and gas entropy obtained by simulating the Santa
  Barbara (SB) cluster \protect\citep{frenk_etal99} with different codes.
  Open and solid triangles are for simulations carried out with the
  ART code \protect\cite{kravtsov_etal02} using $\approx 116,000$ and
  $\approx 928,000$ particles within the virial radius,
  respectively. The solid lines show the corresponding mean profiles
  originally presented in the SB cluster comparison project
  \protect\citep{frenk_etal99}. The open circles show profiles from
  the adaptive mesh refinement (AMR) simulation by Greg Bryan
  performed as part of the SB project. Figure adopyed from
  \protect\cite{kravtsov_etal02} with copyright permission from 2002
  IOP Publishing Ltd.}
\label{sb4}   
\end{figure}

Modern grid-based shock-capturing methods \citep[e.g.,][]{laney_98}
perform well for flows with shocks and contact discontinuities and for
moderately subsonic and turbulent flows. These methods are typically
capable of resolving shocks with just 1-2 grid cells and contact
discontinuities on 3-5 cells.  For given computational resources,
shock-capturing codes can treat far more cells than the SPH codes can
treat particles, but the cells are, in the case of a regular uniform
grid, evenly spaced rather than concentrated into the more interesting
high-density regions as occurs naturally in SPH codes.  This is why in
earlier studies the Eulerian schemes were used mostly for simulations
of large-scale structures and clusters in large volumes
\citep[e.g.,][]{cen_ostriker92, kang_etal94, bryan_norman98,
  miniati_etal00}, while SPH codes were most often used to study
detailed structure of clusters and galaxies simulated in a relatively
small volume \citep[e.g.,][]{evrard90, katz_gunn91, thomas_couchman92,
  katz_white93,evrard_etal94, nfw95, frenk_etal96, yoshikawa_etal00}.
Examples of this are represented by the SPH cosmological simulations
of formation of disk galaxies, which are resolved by several hundreds
of thousands particles \citep[see][ for a
review]{2008arXiv0801.3845M}, and of galaxy clusters, currently
resolved by up to tens of millions gas particles
\citep[e.g.,][]{2008arXiv0808.3401D}.

The situation has changed in the last ten years due to a wider
adoption of adaptive mesh refinement (AMR) techniques for cosmological
simulations
\citep[e.g.,][]{norman_bryan99,abel_etal00,kravtsov_etal02,teyssier02,loken_etal02},
which make shock-capturing gasdynamics schemes competitive with the
SPH. The AMR is used to increase the spatial and temporal resolution
of numerical simulations beyond the limits imposed by the available
hardware. This is done by using fine mesh only in regions of interest
(such as high-density regions or regions of steep gradients in gas
properties) and coarse meshes in other regions (low-density regions or
smooth flows). If only a small fraction of the total volume is
refined, as is commonly done in cosmological simulations, AMR results
in substantial memory and CPU savings compared to the uniform grid of
comparable resolution. The saved computer resources can be used to
dramatically increase resolution where it is really needed. AMR
simulations of the formation of the first stars in the Universe
\citep{abel_etal00}, which have reached a dynamic range in excess of
$10^{10}$, are an impressive illustration of this point.

It is also important to keep in mind that although different numerical
schemes aim to solve the same sets of equations, the actual
discretized versions solved in simulations are not the same. This can
and does lead to differences in the numerical solution
\citep{agertz_etal07,tasker_etal08}. This implies both that different
techniques are useful and complementary in assessing possible
systematic errors associated with a given technique and that such
errors can only be identified in comparisons of simulations performed
using different techniques.

An example of such comparisons is shown in Figure~\ref{sb4}
\citep{kravtsov_etal02}, where results of two independent AMR codes
(which use different techniques for the Poisson and hydrodynamics
solvers) are compared with results of a set of SPH simulations using
the ``Santa Barbara comparison cluster'' \citep{frenk_etal99}. The
figure shows very good general agreement between the two AMR
simulations. Outside the core of the cluster there is also good
agreement between all simulations. However, in the central regions the
entropy profiles of the two Eulerian mesh refinement simulations show
the presence of a well-resolved core, while the mean profile of other
simulations, dominated by the SPH simulations, continue to decrease
monotonically with decreasing radius. The difference is due to
different mixing properties of the codes. The gas in the Eulerian AMR
codes mixes efficiently and the entropy in the core is effectively
homogenized. Mixing in the SPH simulations is absent, however, and the
low entropy gas can survive and sink towards the center via convective
instability \citep{2008arXiv0812.1750M}.  Mixing in SPH simulations
can be increased by either explicitly including a dissipation term in
the equations of hydrodynamics
\citep[e.g.,][]{2008JCoPh.22710040P,2008MNRAS.387..427W} or by
minimizing the degree of artificial viscosity, thereby increasing the
amount of resolved turbulence \citep{dolag_etal05}. The difference in
entropy profiles is also smaller for the "entropy-conservative"
formulation of SPH \citep[see Fig.~1 in][]{ascasibar_etal03}. 
More recently, \cite{springel09} proposed a new hydrodynamic scheme
which is based on an unstructured mesh computed using the Voronoi
tessellation and moving with local fluid flow. Although inherently
Lagrangian, this scheme computes hydrodynamic forces and mass flows
across the mesh boundaries, thus providing the accuracy of Eulerian
codes in describing discontinuities and improving the treatment of
mixing with respect to SPH.

While simple non--radiative simulations require a relatively low
resolution to provide stable predictions about the properties of
the ICM, a reliable description of the processes of star formation,
galaxy evolution and of the feedback effects on the diffuse
inter--galactic and intra--cluster media is far more challenging, both
from a computational and from a physical point of view. Simulations of
cluster formation that aim to describe the effects of galaxy formation
and of energy injection from supernovae and AGNs, include models for
dissipative and heating processes affecting the baryonic
components. These processes are included using phenomenological
parametrization of the relevant physics in the right hand side of
the hydrodynamics equations in the form of additional sink and source
terms
\citep[][]{cen92,cen_ostriker92,katz92,katz_etal96,yepes_etal97}.  For
example, the gas heated by large-scale structure formation shocks (see
\S~\ref{sec:clgrowth} below), supernovae, or AGNs can be allowed to
dissipate its thermal energy via cooling, which is included as a sink
term in the energy equation. The cooling rates are calculated using
local values of gas density, temperature, and (often) metallicity
using either a cooling model of ionized plasma or pre-computed cooling
rate tables \citep[an example of such tables can be found
in][]{sutherland_dopita93}.

Modeling galaxy formation processes in cluster simulations is
extremely challenging because capturing the cosmological environment
within which galaxy clusters form requires sampling scales of few tens
of comoving Mpc, while star formation and relevant gas-dynamical
processes in galaxies occur down to the parsec scales. Therefore,
these processes are modeled using phenomenological "sub-grid" models
describing what occurs at the scales not resolved in simulations.  The
rate at which cold dense gas is converted into collisionless stellar
particles, for example, is parametrized as a power law function of the
local gas density. This rate is used to decrease the gas density and
spawn new collisionless stellar particles with the corresponding mass
\citep[see][for a 
review]{kay_etal02}; dynamics of the stellar particles is then
followed in simulations using the same algorithms as for the
collisionless dynamics of dark matter. It is important to note that
just as dark matter particles in a simulation are meant to sample
characteristics of the collisionless Boltzmann equation and not
represent individual DM particles, stellar particles in simulations
are also samples of the stellar distribution and are not meant
to represent individual stars. Each stellar particle is therefore
treated as a single-age population of stars and its feedback on the
surrounding gas is implemented accordingly.  The feedback here is
meant in a broad sense and, depending on simulations, may include UV
heating, injection of energy and heavy elements via stellar winds and
supernovae, and secular mass loss by stars. Parameterizations of these
processes are motivated by either observations (e.g., SN energy
feedback) or by stellar models (winds, mass loss). Inclusion of
enrichment by heavy elements and tracking of their diffusion allows
one to account for the metallicity dependence of gas cooling rates and
to use metallicity distribution in the ICM as an additional constraint
on the models by comparing it to the observed distributions
\citep[see][ for a review]{2008SSRv..134..379B}.

\section{Cluster formation within a cosmological environment}
\label{sec:nonradprop}

\subsection{How do clusters grow?}
\label{sec:clgrowth}

\begin{figure}
\centerline{
\vbox{
\hbox{
\psfig{file=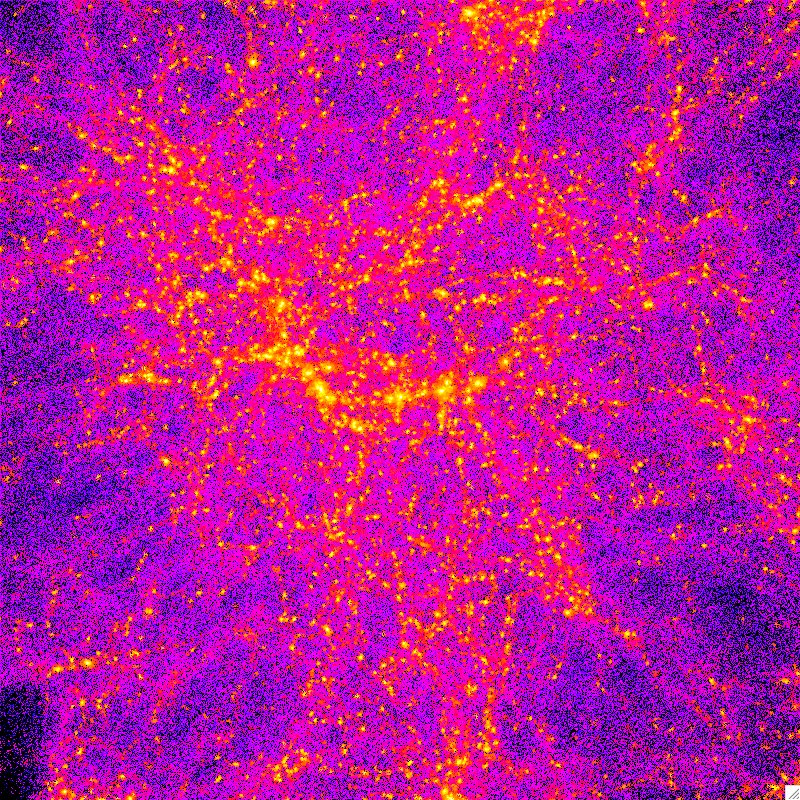,width=5.5truecm}
\psfig{file=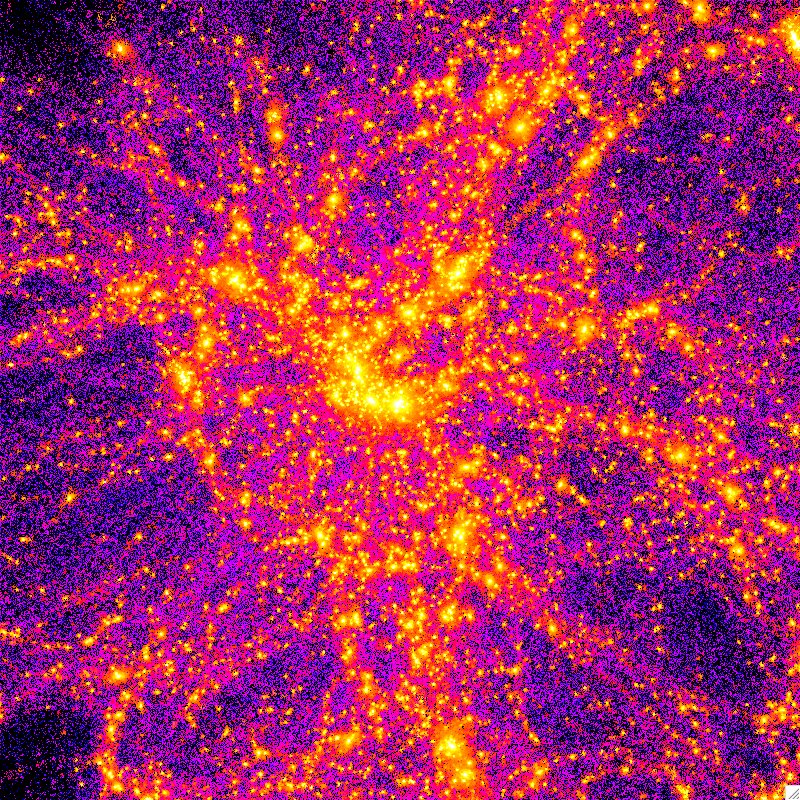,width=5.5truecm}
\psfig{file=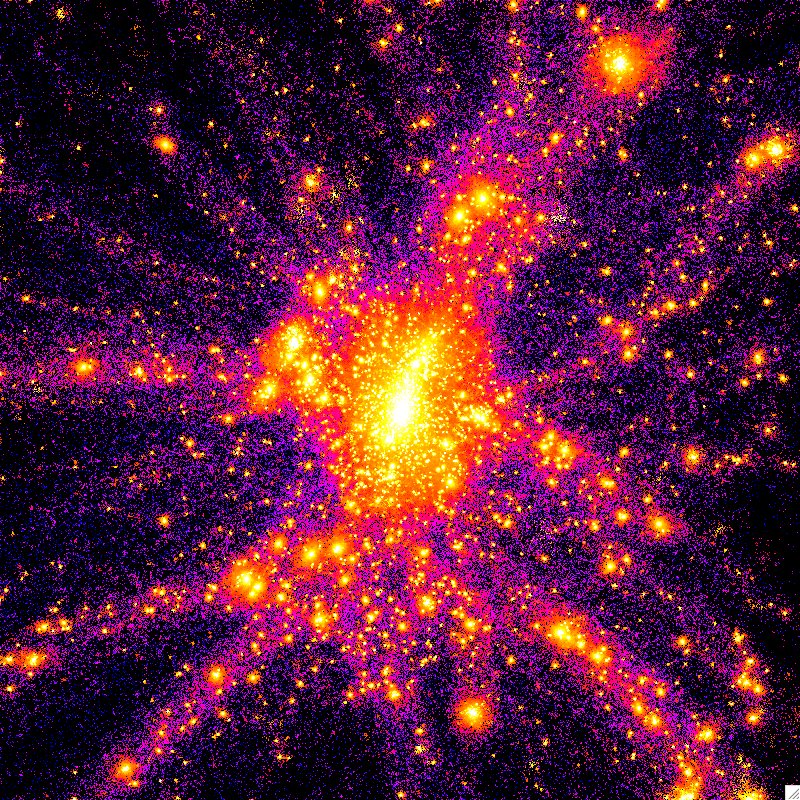,width=5.5truecm}
}
\hbox{
\psfig{file=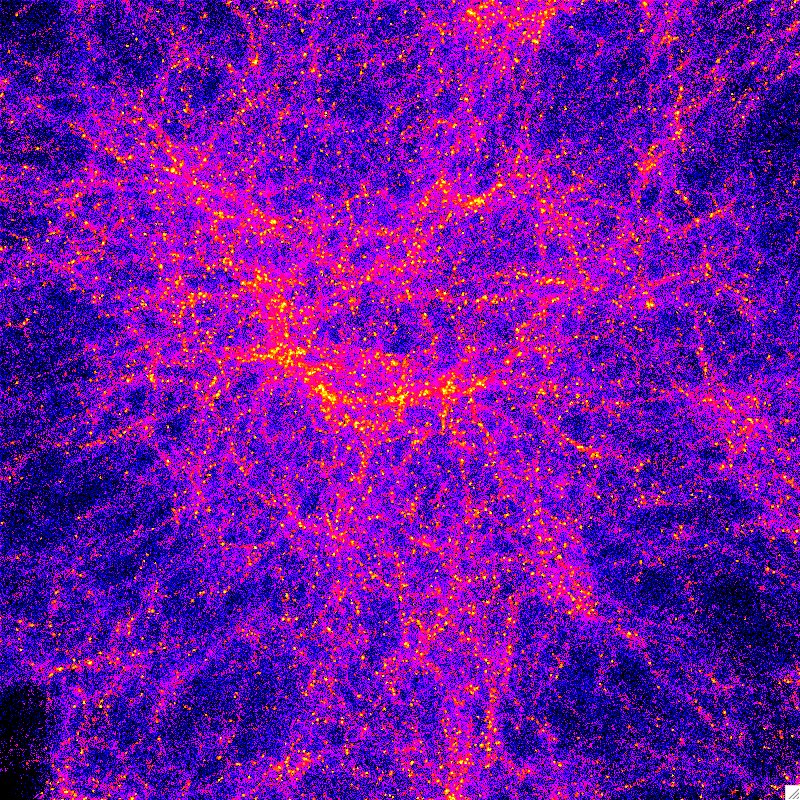,width=5.5truecm}
\psfig{file=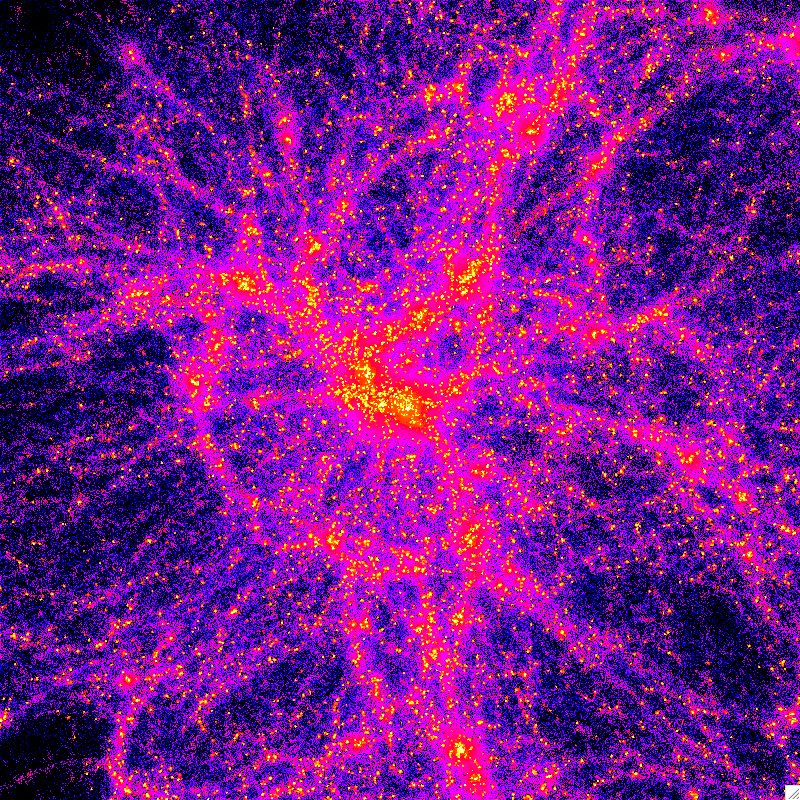,width=5.5truecm}
\psfig{file=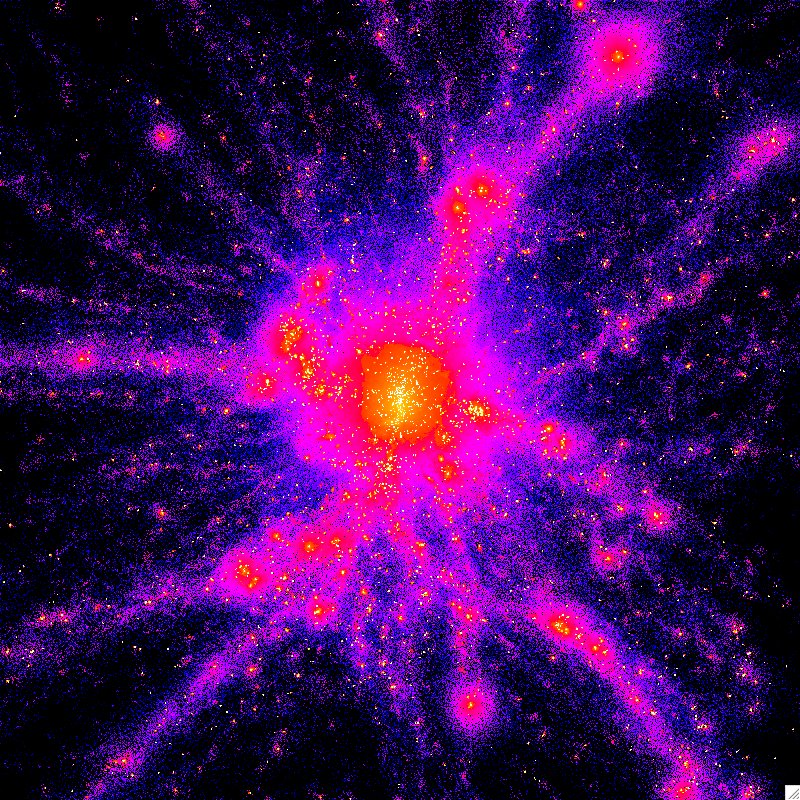,width=5.5truecm}
}
\hbox{
\psfig{file=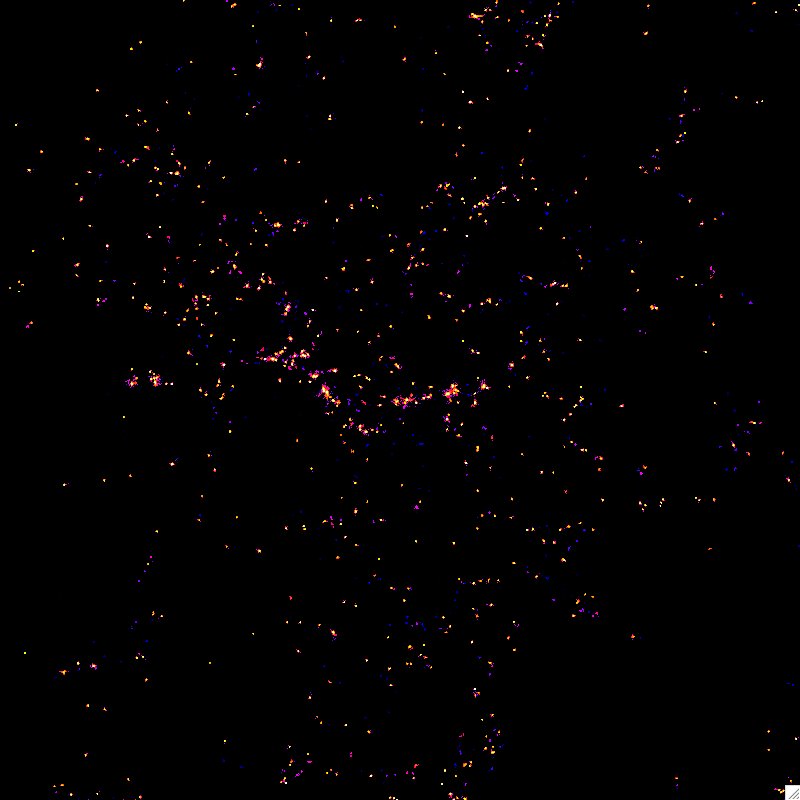,width=5.5truecm}
\psfig{file=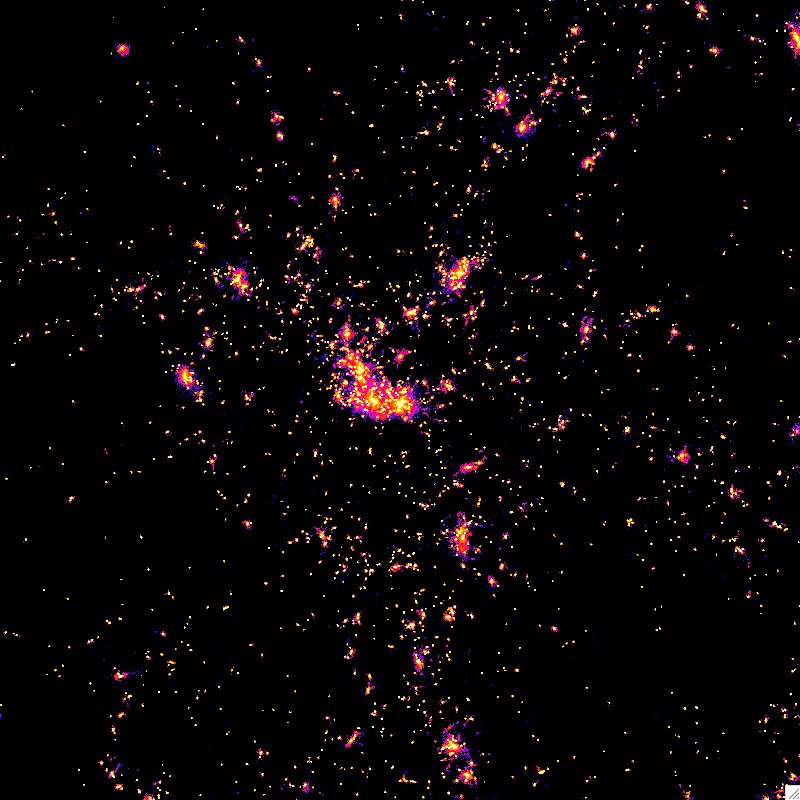,width=5.5truecm}
\psfig{file=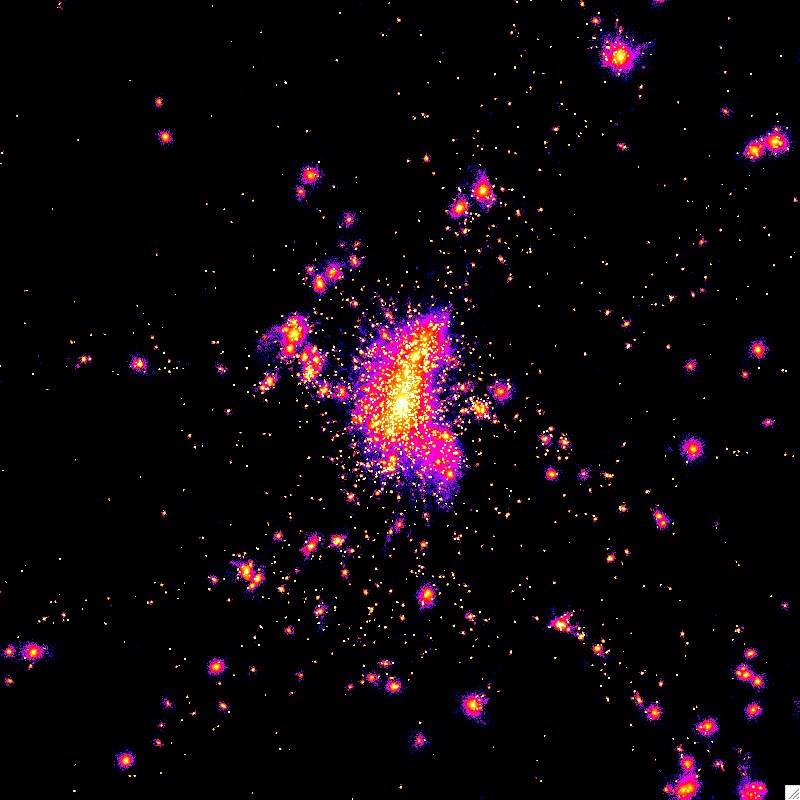,width=5.5truecm}
}
}
}
\caption{The formation of a galaxy cluster in a cosmological context,
  as described by a hydrodynamical simulation carried out with the
  Tree-SPH GADGET code \protect\citep{springel_etal05}. Upper,
  central and bottom panels refer to the density maps of dark matter,
  gas and stellar distributions, respectively. From left to right we show
  the snapshots at $z=4$, where putative proto-cluster regions are
  traced by the observed concentrations of Lyman--break galaxies and
  Lyman--$\alpha$ emitters \protect\citep[e.g.][]{2008ApJ...673..143O}, at
  $z=2$, where highly star--forming radio--galaxies should trace the
  early stage of cluster formation
  \protect\citep{2006ApJ...650L..29M,2009MNRAS.392..795S}, and at
  $z=0$. This cluster has a total virial mass $M_{\rm vir}\simeq
  10^{15}\msun$ at $z=0$ \protect\citep{2008arXiv0808.3401D}. Each
  panel covers a comoving scale of about $24\hm$, while the cluster
  virialized region at $z=0$ is nearly spherical with a radius of
  about $3\hm$.}
\label{fig:mapevol}
\end{figure}

\begin{figure}[t]
\centerline{ 
   \epsfysize=1.5truein  \epsffile{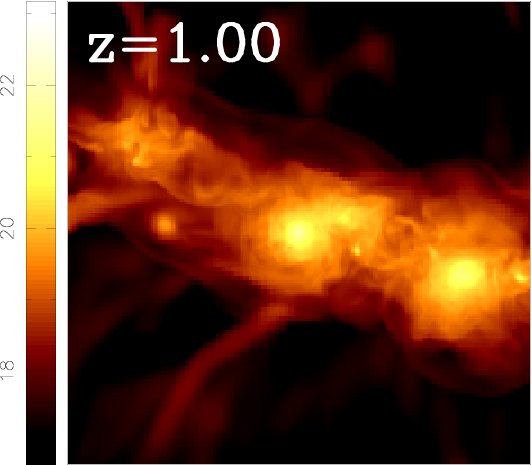}
   \epsfysize=1.5truein  \epsffile{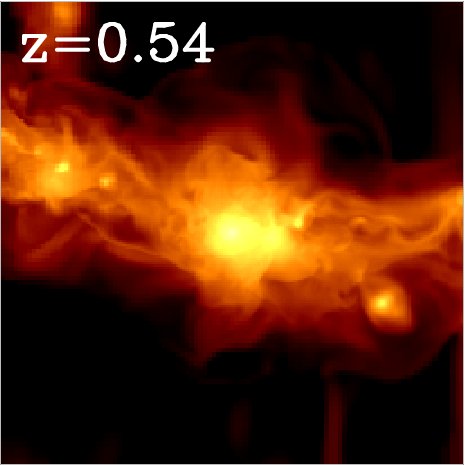}
   \epsfysize=1.5truein  \epsffile{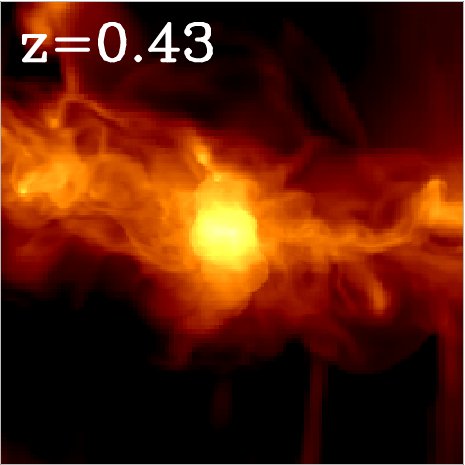}
   \epsfysize=1.5truein  \epsffile{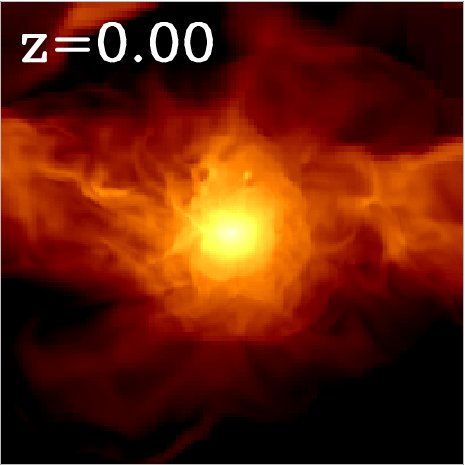}
}                                                  
                                                   
\centerline{ 
   \epsfysize=1.5truein  \epsffile{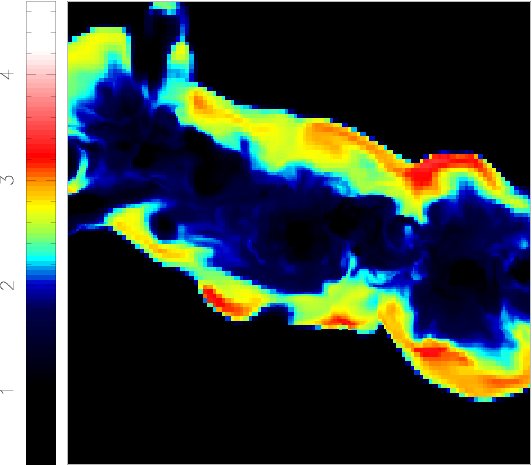}
   \epsfysize=1.5truein  \epsffile{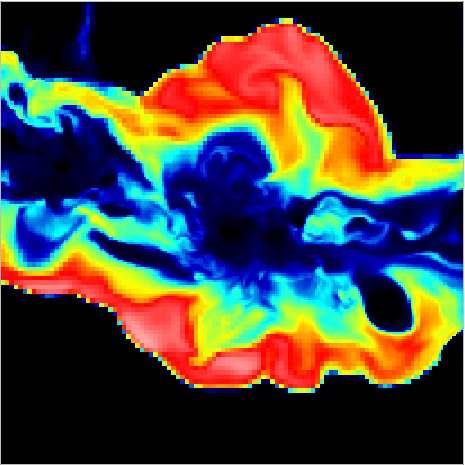}
   \epsfysize=1.5truein  \epsffile{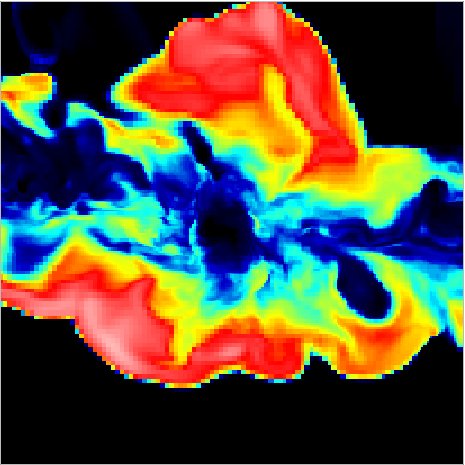}
   \epsfysize=1.5truein  \epsffile{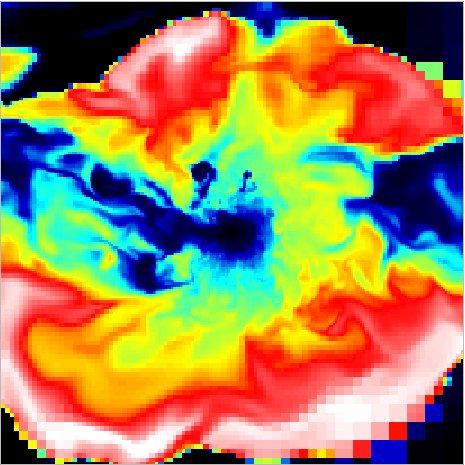}
}
\caption{
The maps of projected density (top) and entropy 
(bottom) of a simulated $\Lambda$CDM cluster at four different redshifts in a
$60h^{-1}$~kpc slice centered on the central density peak. The maps
are color-coded on a $\log_{10}$ scale in units of cm$^{-2}$ (column
density) and keV cm$^2$ (entropy). The size of the region shown is
8$h^{-1}$Mpc. The maps reveal a very complex entropy distribution of
the gas. Both the filaments and the forming cluster are surrounded by
strong accretion shocks.  Note, however, that the accretion shock
around the cluster is very aspherical and does not penetrate into the
filament; relatively low-entropy gas accreting onto cluster along the
filament does not pass through the strong virial shocks and can be
traced all the way to the central $0.5h^{-1}$Mpc of the
cluster. Figure adopted from \protect\cite{nagai_kravtsov02} with copyright
permission from 2005 IOP Publishing Ltd.}
\label{clevol}
\end{figure}

In the hierarchical Cold Dark Matter models of structure formation,
virialized systems of all masses form via a sequence of accretions and
mergers of smaller objects. During this sequence the evolution of each
of the three main components of clusters (dark matter, diffuse gas,
and stars) is deeply interconnected with the others. The DM component
drives the gravitational collapse and the hierarchical accretion of
smaller systems.  To first approximation, gas follows this accretion
pattern. Figure \ref{fig:mapevol} shows how the distributions of DM
(upper panels), gas (central panels) and stars (lower panels) evolve
across cosmic time inside the region forming a cluster, as predicted
by a cosmological hydrodynamical simulation. The gas distribution
generally traces the DM distribution, with its pressure support making
it smoother below the Jeans length scale. Furthermore, stars form
since early epochs within high density halos, where gas can
efficiently cool over a short time scale, thus making their
distribution quite clumpy. These density maps highlight the
hierarchical fashion in which the formation of cosmic structures
proceeds. At early epochs large numbers of small DM halos are already
in place and their distribution traces the nodes of a complex
filamentary structure of the cosmic web. As time goes on,
these filaments keep accreting matter, while small  halos
flow along them, finally merging onto larger halos, placed at the
intersection of filaments, where galaxy clusters form. By the present
time (left panels) a rather massive galaxy cluster has formed at the
intersection of quite large filamentary structures. The virialized region of the cluster
hosts a large number of galaxies which survived mergers of their diffuse
dark matter halos, and is permeated by a nearly spherical gas distribution, 
resembling the observational image shown in Figure \ref{fig:a1689}.

In the case of galaxy clusters, the process of hierarchical merging
and accretion is particularly energetic due to the large masses of
systems involved and filamentary structures that surround them. Strong
gravitational pull of the collapsing cluster region can accelerate
dark matter and gas to thousands of kilometers per second.  Such
high-velocity gas flows are initially strongly supersonic, with Mach
numbers $M\sim 10-1000$, as the gas in the low density regions is
relatively cold
\citep[e.g.][]{miniati_etal00,ryu_etal03,pfrommer_etal06,kang_etal07}.
The supersonic flow then undergoes a shock and converts most of the
acquired kinetic energy into internal thermal energy. These
circumcluster shocks convert tremendous amounts of energy, $\sim
10^{61}-10^{65}$~ergs, which makes cluster formation the most
energetic events since the Big Bang.

These processes are illustrated in Figure~\ref{clevol} \citep[adopted
from][]{nagai_kravtsov02}, which shows maps of projected gas density
(top) and entropy (bottom) around a forming cluster in a $\Lambda$CDM
cosmological simulation at four different redshifts. The figure
illustrates the complex dynamical processes accompanying cluster
formation: accretion of clumps and diffuse gas along filaments, strong
accretion and virial shocks both around cluster and surrounding
filaments, weaker merger shocks within the virial shock of the
cluster, and the complicated flow pattern of the ICM gas revealed by
the entropy map.  Between $z=1$ and $z=0.5$, the main cluster
undergoes a nearly equal mass merger.  More common are smaller mergers
and accretion of groups along large-scale filaments. The mergers play
an important role as they generate shocks which heat the ICM inside
out and are responsible for the bulk of its thermalization
\citep{mccarthy_etal07}. Mergers and accretion flows also generate
turbulent motions and ICM mixing
\citep{norman_bryan99,ricker_sarazin01,nagai_etal03,sunyaev_etal03,dolag_etal05}
and possibly magnetic fields \citep{roettiger_etal99}. Recently, the
shocks and temperature discontinuities accompanying cluster mergers
have been detected in a number of clusters with sensitive X-ray
observations by the {\sl Chandra} satellite \citep[see][for a recent
comprehensive review]{markevitch_vikhlinin07}.

The strong shocks discussed above are apparent as sudden jumps in
entropy (from black to color). The figure shows that accretion shocks
have a non-trivial topology, as they surround both the merging
clusters and the filament along which these clusters move. Note also
that virial shocks do not penetrate into the filament itself and it is
in fact difficult to separate the virial shocks of clusters from the
accretion shocks around the filament.

During the evolution from $z=1$ to $z=0$, as the mass of the main
cluster grows, the radius and entropy gradient of the virial shocks
steadily increase. It is interesting that the virial radius,
physically motivated by spherical collapse models, marks the
transition from the inflow region, dominated by accreting material, to
the relaxed part of the cluster only approximately
\citep{eke_etal98,cuesta_etal08}. The quasi-spherical accretion
shocks, for example, are generally located outside the virial radius
at $\approx (2-3)R_{\rm vir}$ (right bottom panel in
Figure~\ref{clevol}).

Figure~\ref{clevol} shows that strong accretion shocks also exist and
grow around the filament.  Even at the present epoch the gas flowing
along the filament reaches the central cluster regions without passing
through a strong shock near the virial radius. This is clearly
illustrated in the entropy maps in which relatively low-entropy gas
flowing along the filament can be traced all the way to the central
$0.5\hm$ of the cluster. The virial shocks instead propagate along
directions of the steepest pressure gradient into the low-density
voids.  The flow of filamentary gas, with entropy already increased by
the accretion shocks, has considerably smaller Mach numbers (typically
$\sim 1-10$).  When this gas reaches the cluster core it generates
random, slightly supersonic motions in which it dissipates kinetic
energy (see \S~\ref{sec:turbulence} below).

\subsection{Turbulence in the intracluster medium}
\label{sec:turbulence}

The accretion flows described in the previous section and the motions
of groups and galaxies, which perturb the surrounding medium by
gravitational or hydro interactions, can generate substantial
stochastic motions in the ICM.  Energy of the large-scale turbulent
eddies can cascade down to smaller scales resulting in power law
turbulent energy and velocity spectra. In fact, numerical simulations
of cluster formation generally show that subsonic random flows are
ubiquitous even in apparently relaxed clusters
\citep{norman_bryan99,frenk_etal99,nagai_etal03,rasia_etal04,kay_etal04,faltenbacher_etal05,dolag_etal05,rasia_etal06,nagai_etal07a,lau_etal09}.

Turbulence can have several important effects on the ICM.  It can
facilitate mixing of gas at different radii and correspondingly
exchange of thermal energy, diffusion of heavy elements which would
tend to broaden the centrally peaked abundance profile\footnote{A
strongly centrally concentrated metallicity profile can generally be
expected because 1) the central regions of the cluster are strongly
enriched by stars in the central galaxy and in the envelope
surrounding it and 2) the ram pressure stripping of enriched
interstellar medium of cluster galaxies is most effective in the
cluster core where ICM density and galaxy velocities are large.}
\citep{rebusco_etal05}.  Viscous dissipation of turbulent motions can
lead to secular heating of the ICM.  Turbulence is also believed to
maintain and amplify cluster magnetic fields via dynamo processes
\citep{roettiger_etal99,subramanian_etal06}, 
and to contribute to the acceleration of cosmic rays in the ICM
\citep[e.g., ][]{2007MNRAS.378..245B}.

Quite importantly, incomplete thermalization of random gas motions can
bias observational mass estimates of relaxed clusters, which rely on
the assumption of hydrostatic equilibrium \citep[HSE,
e.g.,][]{vikhlinin_etal06,2008A&A...491...71P}.  Given that only
thermal pressure is taken into account in the HSE analysis, the
presence of random gas motions (or any other non-thermal pressure
component) can contribute to the pressure support in clusters and bias
HSE measurements of the total mass profiles \citep{evrard90}.
Analysis of simulated clusters show that up to $\approx$ 10--20 per
cent of pressure support comes from subsonic turbulent motions of gas
\citep{rasia_etal04,rasia_etal06,nagai_etal07a,2008MNRAS.388.1089I}.
A recent comparison of mass from weak lensing and X-ray HSE mass
measurement by \cite{mahdavi_etal08} shows indeed that the HSE mass
are lower by 10-20 per cent compared to lensing mass measurement,
which may indicate the presence of turbulent (or some other
non-thermal) pressure support in clusters.

Note that random motions not only bias the HSE mass estimates but also
lead to a lower ICM temperature for a given total cluster mass, as
part of the pressure support is contributed by the non-thermal
pressure of random motions. The possible existence of these biases
has important implications for the calibration and interpretation of scaling
relations between total mass of clusters and their observable
properties, such as the spectral X-ray temperature or the
Sunyaev-Zeldovich signal. For instance, one can expect that random
motions were even stronger in the past when the accretion and merger
rate of clusters were higher, which would lead to certain evolution of
normalization of the scaling relations. At a given epoch, the
magnitude of the random motions increases with increasing distance
from the cluster center and their effects can therefore be expected to
be much stronger in the outskirts of clusters than in the core.  A
concentration of the total mass density profile derived from the HSE
analysis would then be biased high, as the outer density profile would
be derived to be steeper than it actually is. This may at least
partially explain the high values of cluster concentrations derived
in such analyses \citep{maughan_etal07,buote_etal07}.

How do simulation predictions on the ICM turbulence compare with
observational data? The only tentative observational detection of ICM
turbulence has been obtained from the spectrum of fluctuations in the
pressure map from deep X--ray observations of the Coma cluster
\citep{2004A&A...426..387S}, which is consistent with those expected
for a Kolmogorov spectrum. While future high resolution SZ
observations will probe whether such pressure fluctuations are
ubiquitous in most clusters, an unambiguous detection of stochastic
motions would be obtained from direct measurements of gas velocities
via X-ray spectroscopy \citep{inogamov_sunyaev03}. The high spectral
resolution required makes measurements challenging with the current
X-ray instruments, while high--resolution spectrometers on the next
generation X--ray satellites will open the possibility to infer the
degree of ICM turbulence. As for simulations, the robustness of their
predictions on ICM turbulence relies on their capability to correctly
describe the physical viscosity of the gas, which is in turn
affected by the presence of magnetic fields (see
Sect. \ref{sec:addphysics} below), and to have the numerical and
artificial viscosity under control.

\section{Making cluster simulations more realistic?}
\label{sec:morereal}

Although gravity is the main driver of galaxy clusters evolution, subtle but
important effects on the properties of the ICM are expected from
feedback processes related to star formation and accretion onto
supermassive black holes. First unambiguous signature of these
effects is provided by the detection of heavy elements, the so--called
metals, within the ICM. Such elements are the product of stellar
nucleosynthesis and chemically enrich the ICM at a level of about 1/3
of the solar value \citep[e.g.,][for a
review]{2008SSRv..134..337W}. This implies that gas--dynamical
processes and galactic outflows should have transported and diffused
metals from the interstellar medium, where they are released by SN, to
the inter-galactic medium
\citep[][ for a review]{2008SSRv..134..363S}.

Starting from the late 1980s, the availability of X--ray observations
for statistically representative samples of clusters demonstrated that
predictions of self-similar models are at variance with a number of
observations. For instance, the observed luminosity--temperature
relation
\citep[e.g.,][]{1998ApJ...504...27M,1999MNRAS.305..631A,2004MNRAS.350.1511O,2008arXiv0809.3784P}
is significantly steeper than predicted, $L_X\propto T^\alpha$ with
$\alpha \simeq 2.5$--3 at the scale of clusters and possibly even
steeper for groups. Furthermore, the measured level of gas entropy in
central regions is higher than expected \citep[e.g.,
][]{2003MNRAS.343..331P,sun_etal08}, especially for poor
clusters and groups. Correspondingly, relatively poor systems were
shown to have a relatively lower amount of gas
\citep[e.g.,][]{vikhlinin_etal06,2008A&A...487..431C}, thus violating the basic
assumption on which the self--similar model is based.  This led to the
notion that some physical process, besides gravity, should be
responsible for the lack of self--similarity in the central regions of
clusters and groups.

In the following sections we will review the attempts that have been
undertaken over the last few years to incorporate effects of galaxy formation and of
feedback energy release from supernovae and AGN into cosmological simulations
of cluster formation. We will conclude with
a discussion of the role played by a number of physical processes
which are expected to take place in the ionized ICM plasma.

\subsection{Cooling vs. heating the ICM with stellar feedback}
\label{sec:ch}
The first mechanism introduced to break the ICM
self--similarity is non--gravitational heating
\citep[e.g.,][]{evrard_henry91,kaiser91,2001ApJ...546...63T}.
Early studies have suggested that observations can be explained if
early ($z>1$) pre-heating by galactic winds and/or AGNs injected
significant amount of energy into the ICM \citep[][see
\citeauthor{voit05} \citeyear{voit05} for a
review]{kaiser91,evrard_henry91}. 

This entropy increase prevents the gas from sinking to the
center of DM halos, thereby reducing gas density and X--ray
emissivity. 
If $E_h$ is the extra heating energy per gas particle, then
this effect will be large for small systems, whose virial
temperature is $k_BT\mincir E_h$, while leaving rich clusters with
$k_BT\gg E_h$ almost unaffected. Therefore, we expect smaller systems
to have a relatively lower gas fraction and X--ray luminosity, thus
leading to a steepening of the $L_X$--$T$ relation.

Cluster simulations in which the ICM was heated either via a model for
supernovae-driven winds from cluster galaxies
\citep{metzler_evrard94,metzler_evrard97,2004MNRAS.348.1078B,dave_etal08}
or via an arbitrary pre-heating at an early epoch
\citep[e.g.,][]{navarro_etal95,mohr_evrard97,2001ApJ...555..597B,2001ApJ...559L..71B,2002MNRAS.336..409B,2002MNRAS.336..527M,2006ApJ...649..640M}
appeared to describe the observed X--ray scaling relations
considerably better. Indeed, a suitable choice of extra energy and
heating redshift can be eventually found, which produces the correct
$L_X$--$T$ relation. However, the amount of energy required to heat
the ICM to the desired levels was too large to be provided by
supernovae for a typical initial stellar mass function
\citep{kravtsov_yepes00}, even assuming high efficiency for the
thermalization of supernovae-injected energy. Furthermore, an
undesirable feature of pre--heating is that it creates fairly large
isentropic core, thus at variance with respect to observational data
showing a fairly low entropy level at the center of relaxed clusters
\citep[e.g.,][]{donahue_etal06}. Finally, observations of the
statistical properties of the Ly$\alpha$ absorption lines in quasar
spectra constrain any pre-heating to take place only in high density
regions of the high-redshift intergalactic medium
\citep{2007ApJ...671..136S,borgani_viel09}.

\begin{figure}
  \centerline{ 
\hbox{
\psfig{file=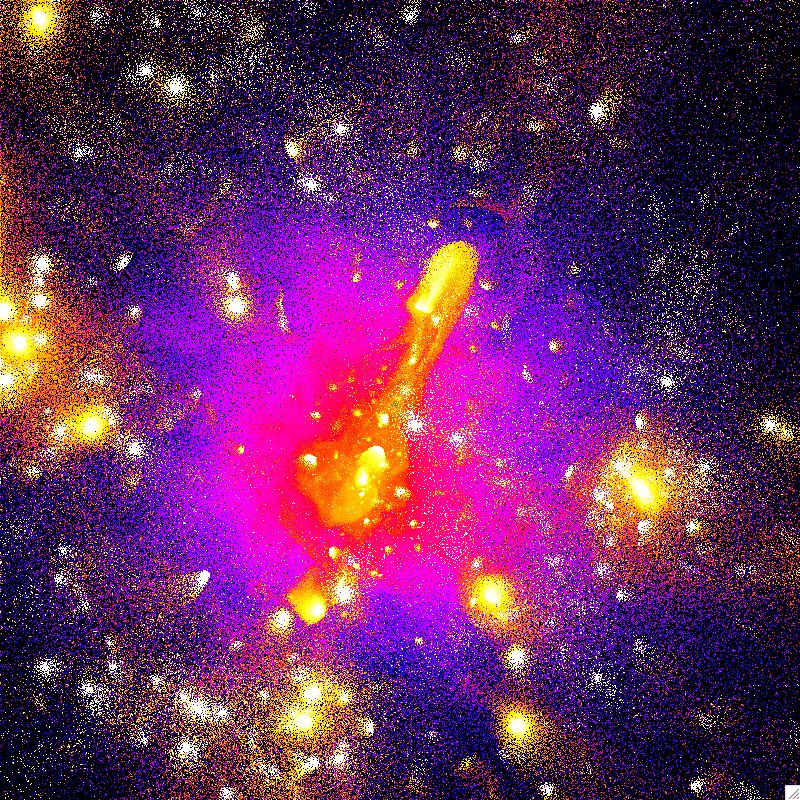,width=8.5truecm} 
\psfig{file=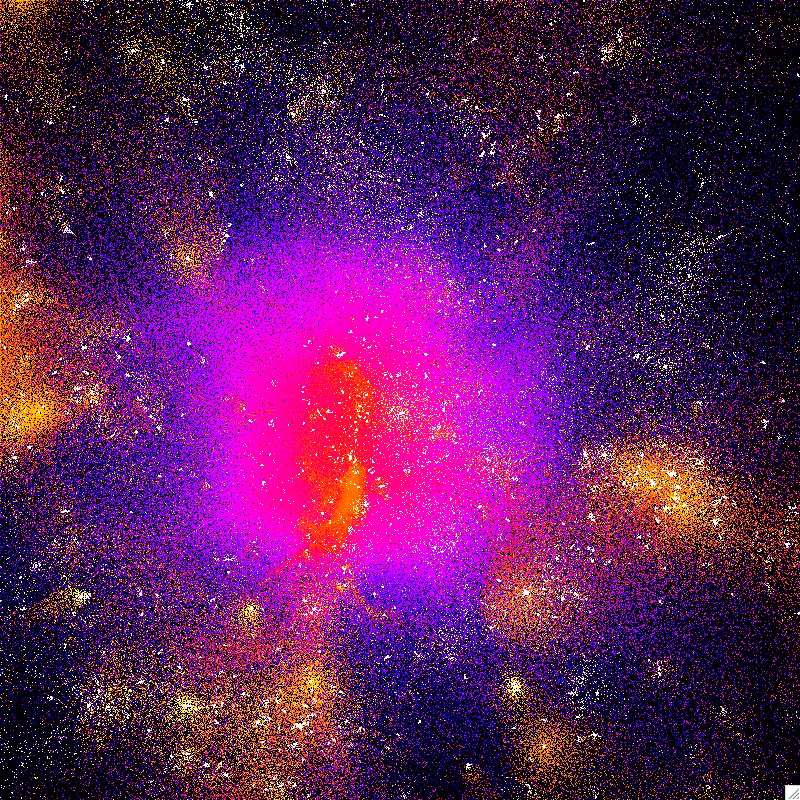,width=8.5truecm} 
}}
\caption{Maps of the gas entropy at $z=0$ for the galaxy cluster shown
  in Figure \protect\ref{fig:mapevol}, for the run with only
  gravitational heating (left panel) and for the run with cooling and
  star formation (right panel). Each panel encompasses a physical
  scale of about 8.5 Mpc. Brighter colors correspond to lower-entropy
  gas, while magenta and blue colors mark gas at progressively higher
  entropy. The simulations have been carried out with the Tree-SPH
  GADGET-2 code \protect\citep{springel_etal05}. The radiative
  simulation also include a description of chemical enrichment and the
  dependence of the cooling function on the gas metallicity
  \protect\citep{2007MNRAS.382.1050T}.}
\label{fig:entr_maps}
\end{figure}

\begin{figure}
  \centerline{ 
\hbox{
\psfig{file=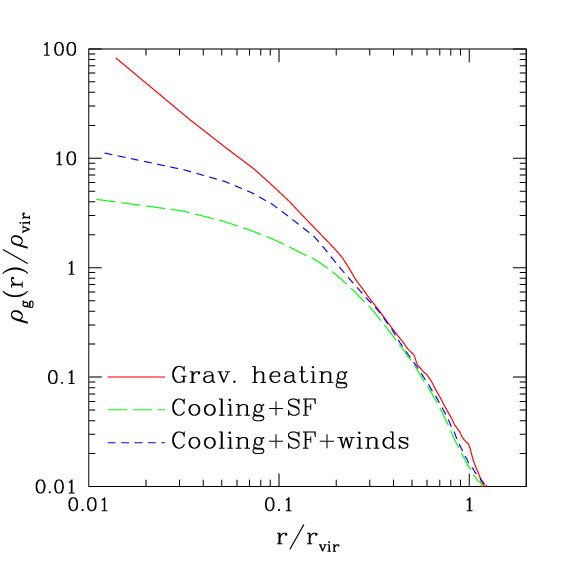,width=8.truecm} 
\psfig{file=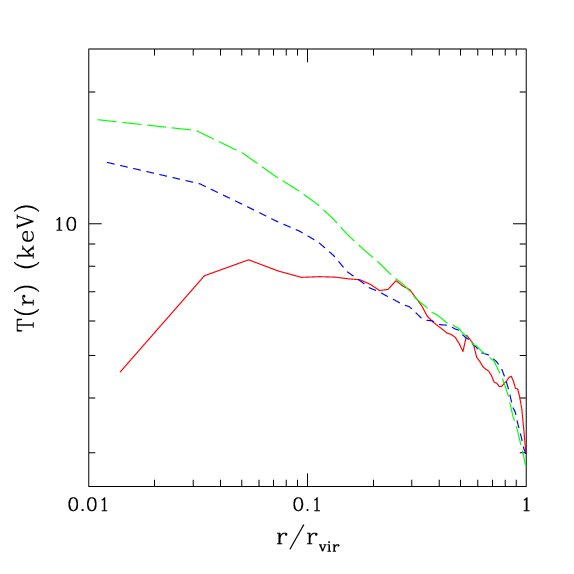,width=8.truecm} 
}}
\caption{Comparison of the gas density (left panel) and temperature
  (right panel) profiles at $z=0$ for the galaxy cluster shown in
  Figure \protect\ref{fig:mapevol}. The three curves are for the run
  with only non--radiative physics (solid red), with cooling and star
  formation (long--dashed green) and also including the effect of
  galactic outflows powered by SN explosions \protect\citep[][
  long--dashed blue]{2003MNRAS.339..289S}.}
\label{fig:profs_rad}
\end{figure}

Although it may look like a paradox, radiative cooling has been also
suggested as a possible alternative to non--gravitational heating to
increase the entropy level of the ICM and to suppress the gas content
in poor systems. As originally suggested by \cite{bryan00} and
\cite{2001Natur.414..425V}, cooling provides a selective removal of
low--entropy gas from the hot phase \citep[see
also][]{2002ApJ...572L..19W}. As a consequence, only gas having a
relatively high entropy will be observed as X--ray emitting. This
analytical prediction is indeed confirmed by radiative hydrodynamical
simulations \citep{pearce_etal00,muanwong_etal01,dave_etal02,valdarnini02,valdarnini03,kay_etal04,nagai_etal07b,ettori_brighenti08}. Figure \ref{fig:entr_maps} shows a comparison between the
entropy maps of the same cluster shown in Figure \ref{fig:mapevol},
simulated by including only gravitational heating (left panel) and
radiative gas cooling along with star formation and chemical
enrichment (which in turn affect the cooling rate; right panel). In
the non--radiative case a number of merging subgroups show up as
clumps of low-entropy gas within a high--entropy atmosphere,
traced by the magenta/blue halo\footnote{In the light of the
  discussion of Sect. \ref{sec:techn}, we expect that these low
  entropy clumps may partly evaporate in the presence of a mixing more
  efficient than that provided by SPH simulations.}. Due to the action
of ram pressure, some of these clumps leave strips of low--entropy gas
behind them, which form comet--like structures. This is quite apparent
for the merging sub-halo located in the upper-right side with respect
to the cluster center, which has just passed the apocenter of its
orbit. This picture drastically changes in the radiative run. In this
case, the removal of low entropy gas in the densest regions, including
the core, has been so efficient that only relatively smooth structures
are visible within the hot cluster atmosphere. A decrease in the
density of the hot gas in central regions, as a consequence of
cooling, is also shown in the left panel of Figure
\ref{fig:profs_rad}, where the gas density profiles in radiative and
non-radiative simulations of this same cluster are compared.

\begin{figure}[t]
\vspace{-0.5cm}
\centerline{ 
  \epsfysize=6truein  \epsffile{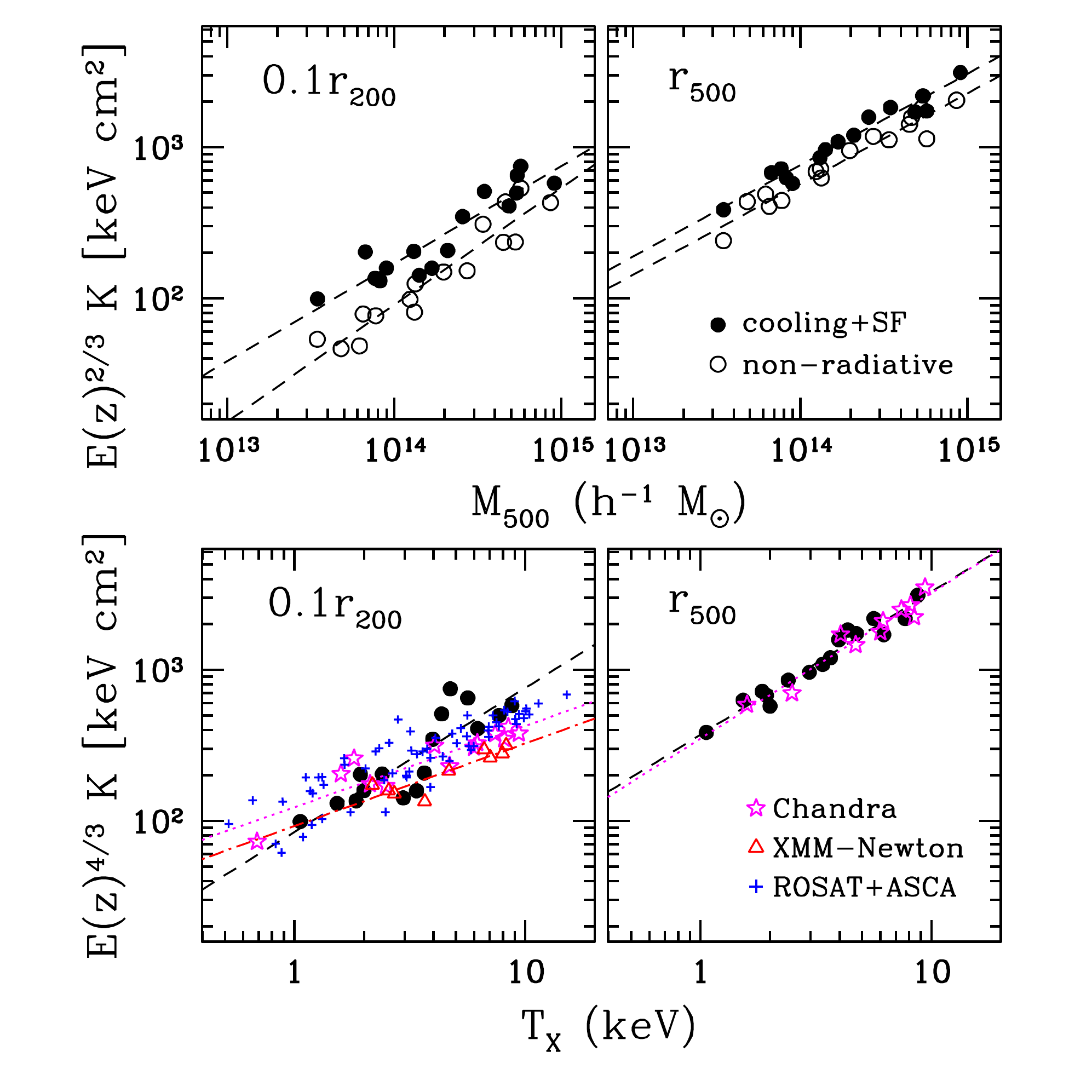}
}
\vspace{-1cm}                                                     
\caption{The panels show scaling of ICM entropy ($k_BT/n_e^{2/3}$,
  where $T$ is the ICM temperature in keV and $n_e$ is its number
  density of electrons in $cm^{-3}$) in the cluster core (0.1 of the
  virial radius, left panels) and at $r_{500}$ (about $0.5r_{200}$,
  right panels). The upper panels show entropy as a function of total
  mass within $r_{500}$ for both non-radiative (open circles)
  simulations and runs with cooling and star formation (solid
  circles). In the simulations with cooling the lowest entropy gas
  cools and condenses out of the hot ICM, which results in the higher
  entropy of the remaining ICM compared to non-radiative
  simulations. The bottom panels show the entropy in simulations with
  cooling (solid circles) compared to different observational
  measurements (as indicated in the legend) as a function of the
  spectroscopic X-ray temperature of the ICM measured excluding the
  inner $0.15r_{r500}$ of the cluster from cluster center as a
  function of the overall X-ray temperature of the ICM. Note that
  self-similar scaling is $K\propto T_{\rm X}$. The simulated points
  are close to the self-similar scaling at both radii. Observations
  exhibit a weak scaling with temperature in the cluster core
  $K\propto T_{\rm X}^{0.5\pm 0.1}$ indicative of additional physical
  processes breaking self-similarity of the ICM. At $r_{500}$,
  however, both observations and simulations with cooling exhibit
  scaling close to the self-similar expectation.  Based on the results
  of \protect\cite{nagai_etal07b}.}
\label{kTx}   
\end{figure}

Note that the dependence of the condensed gas fraction on cluster mass
is weak (see \S~\ref{sec:gasfrac}), thus cooling alone will not be
able to break self--similarity to the observed level (see
Figure~\ref{kTx}).  Furthermore, radiative cooling is a runaway
process and converts too large a fraction of gas into stars. Indeed,
observations indicate that only about 10-15 per cent of the baryon
content of a cluster is in the stellar phase
\citep[e.g.,][]{2001MNRAS.326.1228B,lin_etal03,gonzalez_etal07}. On
the other hand, the radiative simulation of the cluster shown in
Figure \ref{fig:entr_maps} converts into stars about $35\%$ of the
baryons within the virial radius.

Another paradoxical consequence of cooling is that it increases the
ICM temperature at the center of clusters. This is shown in the right
panel of Figure \ref{fig:profs_rad} which compares the temperature
profiles for the radiative and non--radiative simulations of the same
cluster.  The effect of introducing cooling is clearly that of
steepening the temperature profiles at $r\mincir 0.3r_{\rm vir}$.  The
reason for this is that cooling causes a lack of central pressure
support. As a consequence, gas starts flowing in sub-sonically from
more external regions, thereby being heated by adiabatic
compression. This steepening of temperature profiles represents
another undesirable feature of simulations including radiative
cooling.

\begin{figure}
  \centerline{ 
\hbox{
\psfig{file=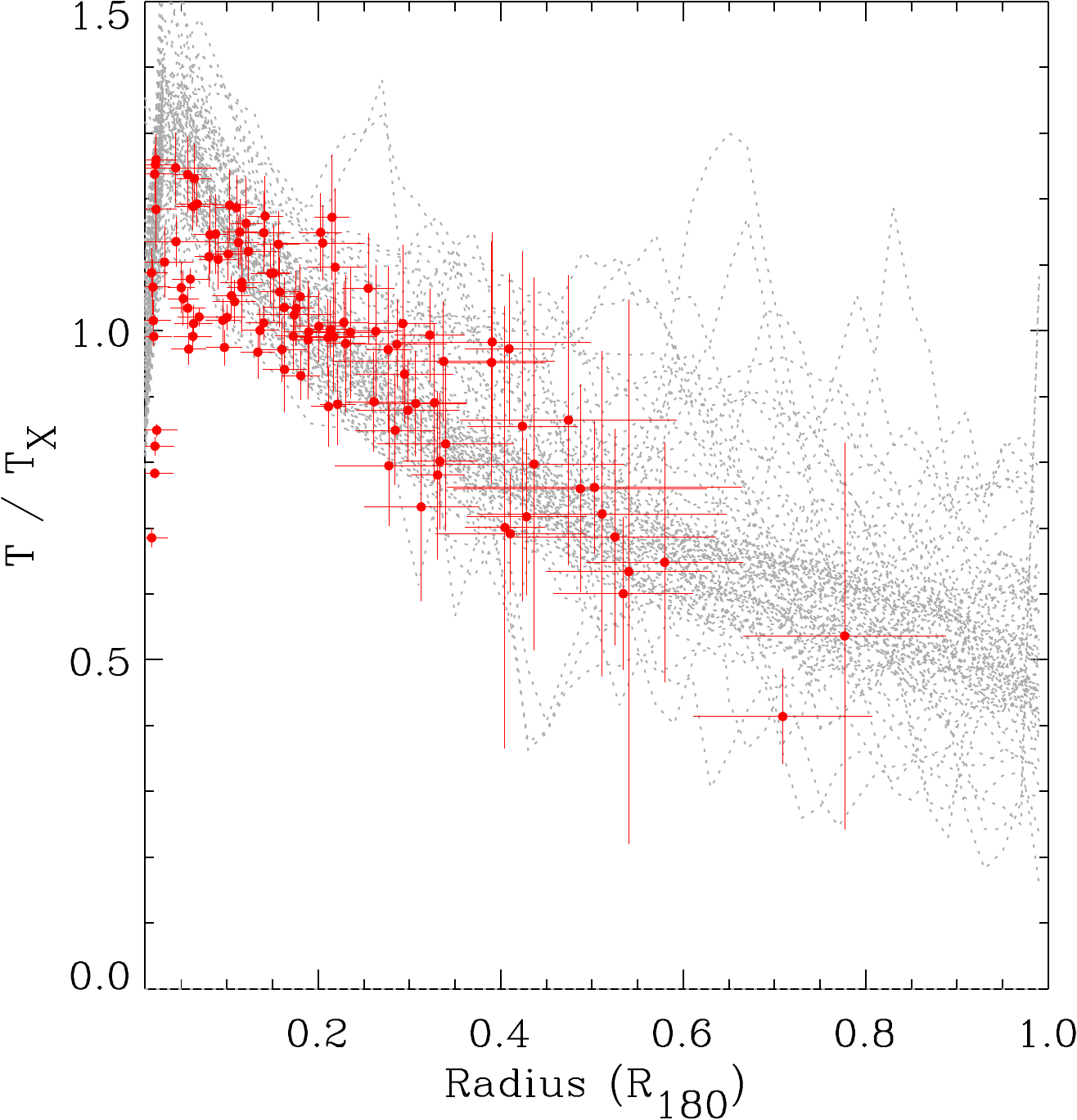,width=7.5truecm} 
\psfig{file=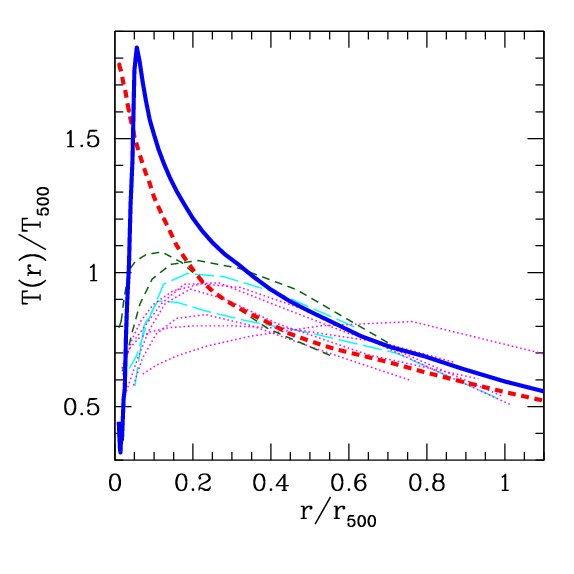,width=8.5truecm} 
}}
\caption{Left panel: comparison between temperature profiles from XMM
  observations of nearby galaxy clusters (errorbar crosses;
  \protect\citealt{2007A&A...461...71P}) and simulations (light
  curves). Figure adopted from \protect\cite{2004MNRAS.348.1078B} with
  copyright permission from 2007 EDP Sciences Ltd. Right panel:
  comparison between simulated and observed temperature profiles. The
  heavy solid curve shows the average profiles from an ensemble of
  clusters simulated including radiative cooling, star formation and a
  prescription for SN feedback. The heavy dashed curve shows the
  corresponding result for non--radiative simulations. Lighter curves
  are the observed temperature profiles for a set of nearby relaxed
  clusters from Chandra data, analyzed by
  \protect\cite{2005ApJ...628..655V}. Figure adopted from
  \protect\cite{nagai_etal07b} with copyright permission from 2007 IOP
  Publishing Ltd.}
\label{fig:tprofs}
\end{figure}

As already discussed in the introduction, cosmological simulations of
clusters are remarkably successful in reproducing the observed
declining temperature profiles outside the core regions
\citep[e.g.,][]{loken_etal02,2006MNRAS.373.1339R}, where gas cooling
is relatively unimportant. This is illustrated in the left panel of
Figure \ref{fig:tprofs}, which shows the comparison temperature
profiles from clusters simulated with the SPH GADGET code
\citep{2004MNRAS.348.1078B} and from a set of clusters observed with
the XMM satellite. Quite apparently, simulated and observed profiles
agree with each other outside the cool core regions. On the other
hand, a number of analyses have demonstrated that radiative
simulations fail at reproducing the observed gentle decline of
temperature profiles in the core regions
\citep[e.g.,][]{2003MNRAS.342.1025T,valdarnini03,nagai_etal07b}.  The
left panel of Figure \ref{fig:tprofs} shows the comparison between
simulated and observed temperature profiles, based on a set of
clusters simulated with the AMR ART code \citep{nagai_etal07b}. This
plot clearly shows that the central profiles of simulated clusters are
far steeper than the observed ones, by a larger amount when cooling
and star formation are turned on.

An eye-ball comparison with the right panel of Figure
\ref{fig:profs_rad} shows a significant difference of the profiles in
the central regions for the non--radiative runs.  While SPH
simulations predict a flattening, or even a decline, of the
temperature profiles at $r\mincir 0.1 r_{vir}$, Eulerian simulations
are characterized by negative temperature gradients down to the
smallest resolved radii. As already discussed in \S~\ref{sec:techn},
this difference is due by the lower degree of mixing and small--scale
dissipation in SPH codes, with respect to Eulerian codes. As expected,
including cooling provides an extra amount of dissipation, thus
reducing the difference between the two hydrodynamical methods.
Clearly, this further calls for the need of performing detailed
comparisons between different simulation codes, for both non-radiative
and radiative simulations, so as to have under control the reason for
these differences.

Steepening of the central temperature profiles and overcooling are two
aspects of the same problem. In principle, its solution should be
provided by a suitable scheme of gas heating which compensates the
radiative losses, pressurizes the gas in the core regions, and 
regulates star formation. Indeed, reproducing the observed star
formation rate represents a challenge not only for cluster
simulations, but more in general for simulations aimed at describing
galaxy formation in a cosmological framework. Energy feedback from
supernova explosions has been originally proposed to generate a
self--regulated star formation. However, the physical processes which
lead to the thermalization of the SN energy clearly require a
sub--resolution description. This is a typical case in which the
detail of the sub--resolution model leads to rather different results
at the resolved scales. In Figure \ref{fig:profs_rad} we also show the
effect of including a kinetic form of energy feedback, in the form of
galactic outflows powered by SN explosions. This feedback mechanism
was originally introduced with the purpose of producing a realistic
cosmic star formation rate \citep{2003MNRAS.339..312S}. Therefore, it
is in principle a good candidate to regulate gas cooling in cluster
cores. The effect of this feedback is to partially compensate
radiative losses and pressurize relatively low-entropy gas, which can
now remain in the hot phase despite its short cooling time. If
compared with the radiative run with no kinetic feedback, the
temperature profile is somewhat flattened, gas density increases by
more than a factor of two in the central regions and the baryonic mass
fraction in stars drops to about 20 per cent.

Although these results go in the right direction, the effect of
stellar feedback is generally not considered the right solution to the
riddle of cool cores.  In the above example, the temperature profile,
although flatter, does not show the gentle decrease in the innermost
region, which would be the signature of a cool core (see the observed
profiles in Figure \ref{fig:tprofs}). Furthermore, star formation rate
in the BCG of this cluster is at a level of about 500
$M_\odot$yr$^{-1}$ \citep{2006MNRAS.373..397S}, thus far larger than
the few tens $M_\odot$yr$^{-1}$ indicated by observations
\citep[e.g.,][]{2006ApJ...652..216R}. As a word of caution, we
  note that current cluster simulations are far from reaching
  high enough resolution to adequately describe the internal structure
  of galaxies and the multi-phase nature of the interstellar
  medium. Indeed, galaxy formation is still an open problem and it is
  not clear whether even simulations of single isolated galaxies, performed at
  much higher resolution \citep[][ for a review]{2008arXiv0801.3845M}, are
  able to produce fully realistic objects. For this reason, the
  approach commonly followed is to explore a range of
  phenomenologically motivated models for star formation and SN
  feedback applied at relatively coarse resolution of cluster 
simulations (scales of $\sim 1-5$~kpc), rather than looking for the values of the parameters
  defining these models, which best reproduce the observational
  properties of galaxy clusters. Much work also remains to be done
to explore sensitivity of the results to increasing resolution and 
inclusion of more sophisticated treatment of interstellar medium 
of galaxies.

More generally, the main observational motivation against stellar
feedback as a mechanism to regulate star formation in cluster cores is
that BCGs have a fairly old stellar population with a low star
formation rate (``read and dead'' galaxies). Therefore, they can not
provide any significant source of feedback from SN explosions. As we
shall discuss in the following, this calls for the need of introducing
some sort of energy feedback, not directly related to star formation
activity, the most popular candidate being represented by AGN
feedback.

\subsection{Modeling AGN heating in cluster simulations}
\label{sec:agn}

Heating from AGN is due to energy released during accretion
of the ICM gas onto a supermassive black hole harbored by the central
cluster galaxy. Effects of AGN activity on the ICM are observed in
many clusters \citep[see][for a review]{mcnamara_nulsen07} and the
energy estimated to be injected by AGNs is sufficiently large to
offset cooling and affect cluster gas in the core and beyond.

In particular, the AGN heating is thought to be the explanation for
the puzzling absence of gas cooling in the central regions of
dynamically relaxed clusters, the so-called ``cool core'' clusters. These
clusters are characterized by a significant excess of X--ray emissivity due
to the high density of gas in the center, which implies rapid
radiative loss of thermal energy at such a high rate  
that all the gas in central cluster regions should cool to low
temperature, $\sim 10^4$K, within a fraction of the typical clusters'
age \citep[e.g.,][]{fabian94}.  Yet, high-resolution X-ray spectra
of core regions obtained during the past decade with the {\sl
XMM}-Newton and {\sl Chandra} space telescopes do not show
significant amount of gas at temperatures below $\sim 0.1-0.3$ of the
virial temperature \citep[see][for a recent
review]{peterson_fabian06}. This implies the presence of steady heating
within the cluster cores. AGN feedback is considered to be the most
likely source of heating, due both to its central location and its
ability to provide sufficient amounts of energy. The AGN heating is
also thought to play an important role in quenching star formation
in the brightest cluster galaxies, thereby reducing the cluster stellar
mass fractions (see \S~\ref{sec:gasfrac}). Trends of hot gas fractions
and entropy in the inner regions with cluster mass also suggest AGN
heating \citep[e.g.,][]{sun_etal08}.

All of this shows the paramount importance of including the AGN
heating in cluster simulations. The challenge, however, is that
details of the heating mechanism are poorly understood. Some models
show that episodic heating in the form of jets and bubbles does not
result in a stable balance between cooling and heating, so that gas
cooling is only delayed but not prevented
\citep{vernaleo_reynolds06}. This reflects a general difficulty to
balance cooling or to prevent cooling instability. First, heating via
an episodic jet may not be available in the specific region where it
is needed to offset the cooling, as a collimated jet can punch a
narrow tunnel through the ICM gas and deposit its energy well away
from the center. Second, even if the heat is distributed by some
process, it does not guarantee stable balance between cooling and
heating because cooling depends on the square of gas density, while
heating generally depends on volume. Finally, the heating process,
while offsetting cooling, should not increase the
entropy significantly, as very low entropy gas is observed in the
central regions of the cool core clusters \citep{donahue_etal06}. 
The presence of such low entropy gas in core regions implies that
this gas needs to be kept continuously pressurized by heating, so that
it is prevented from cooling out of the hot X--ray emitting phase
despite its formally short cooling time. The challenge is how to reach
such a cooling/heating balance in a self-regulated way, without
fine tuning the AGN duty cycle and the efficiency of energy thermalization.

Moreover, while entropy at small radii is observed to be
surprisingly low, at larger radii ($r\sim r_{500}$) it is larger
than predicted by non-radiative simulations
\citep{voit_etal05,sun_etal08}. 
Semi-analytical models predict that the elevated magnitude of entropy
could in principle be explained by smoothing of the intergalactic
medium resulting from strong heating at high redshifts
\citep{voit_ponman03}. However, it is not clear whether such
predictions are confirmed by pre-heated simulations which also include
radiative cooling \citep{borgani_etal05,2007ApJ...666..647Y} or
whether they could also explain the observed nearly self-similar
correlation of entropy at $r_{500}$ with the overall cluster
temperature \citep{nagai_etal07b}.
Heating by both AGN and galactic winds
in general is not expected to scale linearly with cluster mass because
AGN heating depends on the thermal processes and black hole mass on
very small scales, while galactic winds heating should be a function of
total stellar mass in the cluster, which does not scale linearly with
cluster mass.

Keeping these challenges in mind, it is still clear that AGN
heating in some form needs to be included in simulations of cluster
formation for them to reproduce the observed properties of cluster
core regions and ICM properties in smaller clusters and groups of
galaxies. During the last several years, significant theoretical
effort was dedicated to developing models of AGN heating of the ICM
\citep{churazov_etal01,quilis_etal01,ruszkowski_begelman02,brighenti_mathews03,bruggen_kaiser02,omma_binney04,omma_etal04,ruszkowski_etal04,reynolds_etal05,vernaleo_reynolds06,2007MNRAS.376.1547C}. All
of these models, however, considered AGN activity outside the
cosmological context of cluster formation. Although justified by the
short time scale of a single heating episode, 
$\sim 10^8$ yr, one
expects that cool core should be established at rather high redshift,
$z\magcir 2$, when the inter-galactic medium starts reaching high
enough densities around forming proto-BCGs.
The overall effect of many different episodes both in the center of
the cluster and in all of its progenitors can only be estimated by
incorporating the AGN feedback in a full cosmological simulation.

First such simulations have recently been carried out
\citep{sijacki_etal07,2008MNRAS.389...34B,sijacki_etal08}.  Based on
the model for the SMBH formation and growth by \cite{dimatteo_etal05},
\cite{sijacki_etal07} developed a model for AGN feedback via hot
thermal bubbles in a full cosmological simulation of cluster formation
(see also \citealt{2008ApJ...687L..53P}). 
Several of the
results (e.g., the relation between X--ray luminosity and temperature
and the stellar fractions) from the simulations with AGN heating show
a substantial improvement.
However, this heating produces central entropy
profiles inconsistent in both amplitude and shape with observed profiles
exhibiting very low entropy in the inner 100 kpc 
\citep{donahue_etal06}. More recently, \citet{sijacki_etal08}
have extended the model to include the injection of relativistic
cosmic rays (CRs) in the AGN-blown bubbles. Interaction of cosmic rays
with ICM provides a different buoyancy of the bubbles and a more
steady source of heating as their dissipation rates are slow. As a
result, CR heating can successfully regulate cooling flow and can
significantly reduce stellar fraction in clusters.  However, the
resulting entropy profiles are still inconsistent with observations.
This indicates that further improvements to the model will be needed
to reproduce detailed properties of ICM faithfully. Quite
interestingly, \citet{sijacki_etal08} also found that the non--thermal
pressure support associated to cosmic rays can be as large as 50 per
cent at the cluster center. This additional non-thermal pressure
allows the gas to have declining temperature profiles towards the
center without actually cooling.  These results, as well as results of
the CR effects in more idealized models of clusters
\citep{boehringer_morfill88,rephaeli_silk95,mathews_brighenti07,guo_oh08,ruszkowski_etal08,2008ApJ...685..105R},
indicate that cosmic rays are likely an important component of the ICM
physics.

The presence of relativistic electrons is also suggested as a likely
explanation for the observed soft \citep[e.g.,][]{werner_etal07} and
hard \citep[e.g.,][]{sanders_fabian07} excess in the core of few
clusters. In particular, \cite{sanders_fabian07} concluded from
Chandra data that about 40 per cent of the electron pressure at the center of
the Perseus cluster may have non-thermal origin. However, this
conclusion has not been confirmed by the XMM observations by
\cite{2009A&A...493...13M}, who placed instead stringent upper limits
to any hard excess in the X--ray spectrum from the core of
Perseus. Furthermore, \cite{churazov_etal08} derived an upper limit of
10--20 per cent to the non-thermal pressure in the core regions of two
clusters, by combining X--ray and optical data. There is no doubt that
the combination of observations at radio
\citep[e.g.,][]{2008A&A...486L..31C}, optical, X--ray and
$\gamma$--ray \citep{2006ApJ...644..148P,aharonian_hessclus08}
frequencies will allow us to understand the role played by cosmic rays
play in cluster cores and, therefore, whether they need to be
carefully modeled in cosmological simulations of clusters.

\subsection{Modeling other physical processes in the intracluster plasma}
\label{sec:addphysics}

Althoug almost all cosmological simulations discussed in the previous
section treat intracluster plasma as an ideal inviscid fluid, the
actual detailed processes in the ICM can be quite a bit more complex. On
the one hand, the Debye screening length in the ICM is short
($\lambda_{\rm D}\sim 10^7$~cm) and the number of particles within
$\lambda_{\rm D}$ is large. The plasma can thus be treated as a
neutral, continuous field on the scales resolved in cosmological
simulations ($>$~kpc$=3.0856\times 10^{21}$~cm). On the other hand,
the weakness of the Coulomb interactions means that, in the absence of
magnetic field, the mean free path of the electrons is quite large:
$\lambda_e\approx 23\,{\rm kpc}(T_e/10^8\,{\rm K})^2(n_e/10^{-3}\,{\rm
  cm^{-3}})^{-1}$ \citep[e.g.,][]{spitzer62,sarazin86}. For the
typical conditions in the cluster outskirts, $T\sim {\rm few}\times
10^7$~K and $n_e\sim 10^{-4}\,\rm cm^{-3}$ at $r_{500}$
\citep[e.g.,][]{vikhlinin_etal06}, this mean free path can be as large
as $\sim 100$~kpc, and the plasma formally cannot be considered as a
strongly collisional gas on the scales resolved in simulations.

One can then ask whether the hydrodynamic treatment is at all
reasonable. The answer is ``yes'' for two reasons. First, even a
small, dynamically unimportant magnetic field can shorten the mean
free path of the electrons significantly, at least in the direction
orthogonal to the field line. Magnetic fields are indeed observed in
cluster cores \citep[e.g.,][]{eilek_owen02,carilli_taylor02} and there
is a body of indirect evidence of their existence throughout the
cluster volume
\citep[e.g.,][]{carilli_taylor02,govoni_feretti04,eilek_etal06,markevitch_vikhlinin07}.
In addition, magnetic fields are also predicted to be effectively
amplified from the primordial values in mergers during cluster
formation
\citep{roettiger_etal99,dolag_etal99,dolag_etal02,schindler02,bruggen_hoeft06,subramanian_etal06,dolag_stasyszyn08}.
The questions are only how dynamically important the magnetic fields
actually are, the actual magnitude of the mean free path suppression
(which depends on the topology of the magnetic fields, see related
discussion of conductivity below), and whether magneto-hydrodynamic
effects are important in the ICM.  In fact, MHD cosmological
simulations suggests that magnetic fields generally provide a minor
contribution to the total pressure \citep{2001A&A...369...36D}, while
they can become dynamically non-negligible in peculiar cases involving
strong mergers \citep{dolag_schindler00} or in strong cooling flows
\citep{dubois_teyssier08}.

Second, from plasma experiments (as well as observations of supernova
remnants) it is clear that real plasmas exhibit a variety of
excitation modes, in which electrons and ions move collectively in a
correlated fashion. Fluctuations of the electromagnetic field
generated by such motions appear to be much more efficient in
scattering individual electrons than the Coulomb interactions. This
means that the actual mean free path can be much smaller than the
value for the Coulomb scatterings in unmagnetized plasma, at least in
regions such as shocks, where deviations from local thermodynamic
equilibrium are significant. This and the expected presence of
magnetic fields probably make the use of hydrodynamic treatment of gas
in cluster simulations justified.

Nevertheless, it is definitely worth investigating possible effects and
ramifications of the breakdown of hydrodynamic limit. 
One approach is to treat the gas
as a continuous fluid, but with kinetic perturbative correction
terms to be added to the Vlasov equation, so as to account for the
``rarefied'' nature of the gas \citep{2003PhFl...15.3558S}. 
The collisionless effects and deviations from the collisional 
equilibrium can also be treated via explicit modeling of separate
kinetics of electrons and ions \citep[][]{chieze_etal98,teyssier_etal98,takizawa99,2005ApJ...618L..91Y}.

Whether the ICM can be treated as an inviscid fluid, as is done in
most cosmological simulations of clusters, is a different
question. Transport processes can be efficient and important for
thermodynamics in clusters \citep[see, e.g.,][for a
review]{sarazin86}.  Some amount of viscosity is introduced in the
codes for numerical reasons, but this {\it artificial} viscosity,
although important by itself \citep[see, e.g.,][for detailed
discussion]{dolag_etal05}, should be distinguished from the real
physical viscosity. The latter, however, is another factor that can
affect important processes in the ICM, such as ram pressure stripping,
dissipation of sound waves and random gas motions, survival of AGN
bubbles.  \citet{sijacki_springel06} have investigated the effect of
adding the Spitzer--Braginskii physical viscosity (in addition to the
numerical artificial viscosity), by implementing an SPH formulation of
the Navier-Stokes equation. They find that the contribution of the
physical viscosity is important and results in additional source of
entropy due to viscous dissipation of motions generated by mergers and
accretion accompanying cluster formation. Adding physical viscosity
also makes ram pressure stripping of gas from infalling groups and
galaxies more efficient \citep[although see][]{roediger_bruggen08} and
changes the morphology and disruption of the AGN-inflated
bubbles. Since this viscosity is related to the Coulomb mean free path
of the ions, it can be efficiently suppressed by the presence of
magnetic fields. Therefore, its detailed description requires in
principle a self--consistent description of the intra-cluster magnetic
field.

Heat conduction is another important transport process to consider, as
was pointed out by a number of studies over the last twenty years
\citep[e.g.,][]{tucker_rosner83}, if the conductivity is close to the
\citet{spitzer62} value for unmagnetized plasma. This is because the
ICM is not isothermal and temperature gradients exist both in the core
and in the outer regions (see Figure~\ref{fig:tprofs}). In the context
of the cooling flow problem, efficient conduction could provide a way
to tap a vast reservoir of thermal energy at large radii in
clusters. The value of the actual conductivity is highly uncertain due
to uncertainties in the plasma processes and topology of intracluster
magnetic fields
\citep{tribble89,chandran_cowley98,narayan_medvedev01,malyshkin01}.
An efficient conduction tends to create an isothermal core
\citep[e.g.,][]{dolag_etal04}.  However, even when the conduction is
close to the maximum Spitzer value, it may not be sufficient to fix
all the problems in cluster modeling by itself. In particular, due to
strong expected temperature dependence of conductivity (e.g, the
Spitzer conductivity $\kappa_S$ scales with temperature as
$\kappa_S\propto T^{5/2}$) it is impossible to offset radiative
cooling in low mass clusters, even if the conductivity is sufficient
to offset cooling in high mass clusters
\citep{voigt_fabian04,dolag_etal04}.  Furthermore, conduction enters
in a saturation regime whenever the typical scale of a temperature
gradient falls below the electron mean free path. Therefore, whenever
radiative cooling takes over, conduction may not be efficient enough
to compensate radiative losses by a heat transfer across the interface
between cold and hot gas phases. This is the reason why conduction
has a minor effect in decreasing the stellar mass fraction in
cluster simulations \citep{dolag_etal04}.  Conduction also requires
fine-tuning \citep{bregman_david88,conroy_ostriker08} and the thermal
equilibrium mediated by conduction tends to be unstable
\citep{kim_narayan03}, although the problem may be mitigated by the
regulation of conductivity by MHD turbulence related to
magneto-thermal instabilities \citep{balbus_reynolds08}. It is
interesting to note that heat transport can operate not only as a
kinetic effect via plasma particle scatterings, possibly suppressed
by the presence of magnetic fields, but also as a heat carried over
significant range of radii by large-scale subsonic turbulent eddies
\citep{cho_lazarian04}. In this case, the conduction can be quite
efficient and comparable to the Spitzer conductivity and would not
sensitively depend on the presence and topology of magnetic fields.

A potentially important heating mechanism in the ICM could be wakes
generated by supersonic galaxy motions \citep{elzant_etal04}, although
this process alone may not solve the cooling flow problem
\citep{kim_etal05,kim07,conroy_ostriker08}.  Motions by galaxies and
infalling groups can also be an important driver of turbulent motions
\citep[e.g.,][]{kim07} that affect transport processes and can
viscously dissipate their energy thereby heating the ICM
\citep{dennis_chandran05}.  Although galaxy motions are modeled in
high-resolution hydrodynamics simulations of clusters
\citep{faltenbacher_etal05,nagai_kravtsov05,maccio_etal06}, the
current resolution may not be sufficient to follow the formation of
wakes and transport of energy properly. Further simulations of this
process for realistic clusters are needed.

Due to space limitation, this is but a brief overview of the examples
of physical processes that can influence various aspects of the ICM
evolution, but are often neglected in cosmological simulations of
cluster formation. A more in depth discussion of these processes can
be found in a recent review by \cite{dolag_etal08}. The main purpose
of our overview here was to give a sense of the effects and, most
importantly, to highlight the fact that some aspects of the ICM
physics are still rather poorly understood. This means that as many
checks of models against observations as possible are required to make
sure the models are a reasonable description of reality. We discuss
some examples of how to constrain the ICM physics using observations
below.

\subsection{Constraints from the baryon and gas fractions in clusters}
\label{sec:gasfrac}
Measuring the fraction of the total mass in baryons within clusters,
$f_b$, is one of the most powerful means to measure the density
parameters contributed by matter, $\Omega_m$, and by dark energy. Once
the value of the baryon density parameter, $\Omega_b$, is known
(e.g. from observations of the CMB anisotropies or from primordial
nucleosynthesis arguments), then measuring $f_b$ for nearby clusters
turns into a measurements of the matter density parameter, owing to
$\Omega_m=\Omega_b/f_b$
\citep[e.g.,][]{whitetal93:baryoncontent,1995MNRAS.273...72W,1995ApJ...445..578D,1999MNRAS.305..834E,2000A&A...361..429R,2003A&A...403..433C}.
This method is based on the assumption that clusters contain a cosmic
share of baryons. Furthermore, since the gas fraction measured from
X--ray observations depends on the luminosity distance at the cluster
redshift, assuming that this fraction does not evolve turns into a
geometrical cosmological test, which allows one to also constrain the
dark energy content of the universe
\citep[e.g.,][]{1996PASJ...48L.119S,pen97,2002MNRAS.334L..11A,2003A&A...398..879E,laroque_etal06,allen_etal08}. Owing
to the previous discussion on the effects of cooling and heating 
on the distribution of baryons within
clusters, one may wonder how well funded are the above
assumptions. For instance, while cooling results in the condensation
of baryons, heating prevents gas from sinking in the cluster potential
wells, thus potentially changing the value of $f_b$ in the cluster
environment. Furthermore, since both gas cooling and heating rates are
expected to change with time, the value of $f_b$ may also evolve with time.
While the baryon fraction test has provided convincing constraints on
the value of $\Omega_m$, precision cosmology based on $f_b$
measurements for distant cluster requires the above uncertainties to
be controlled with good precision, even assuming that the total
cluster mass can be perfectly known.

\begin{figure}
\centerline{  \epsfysize=7truein  \epsffile{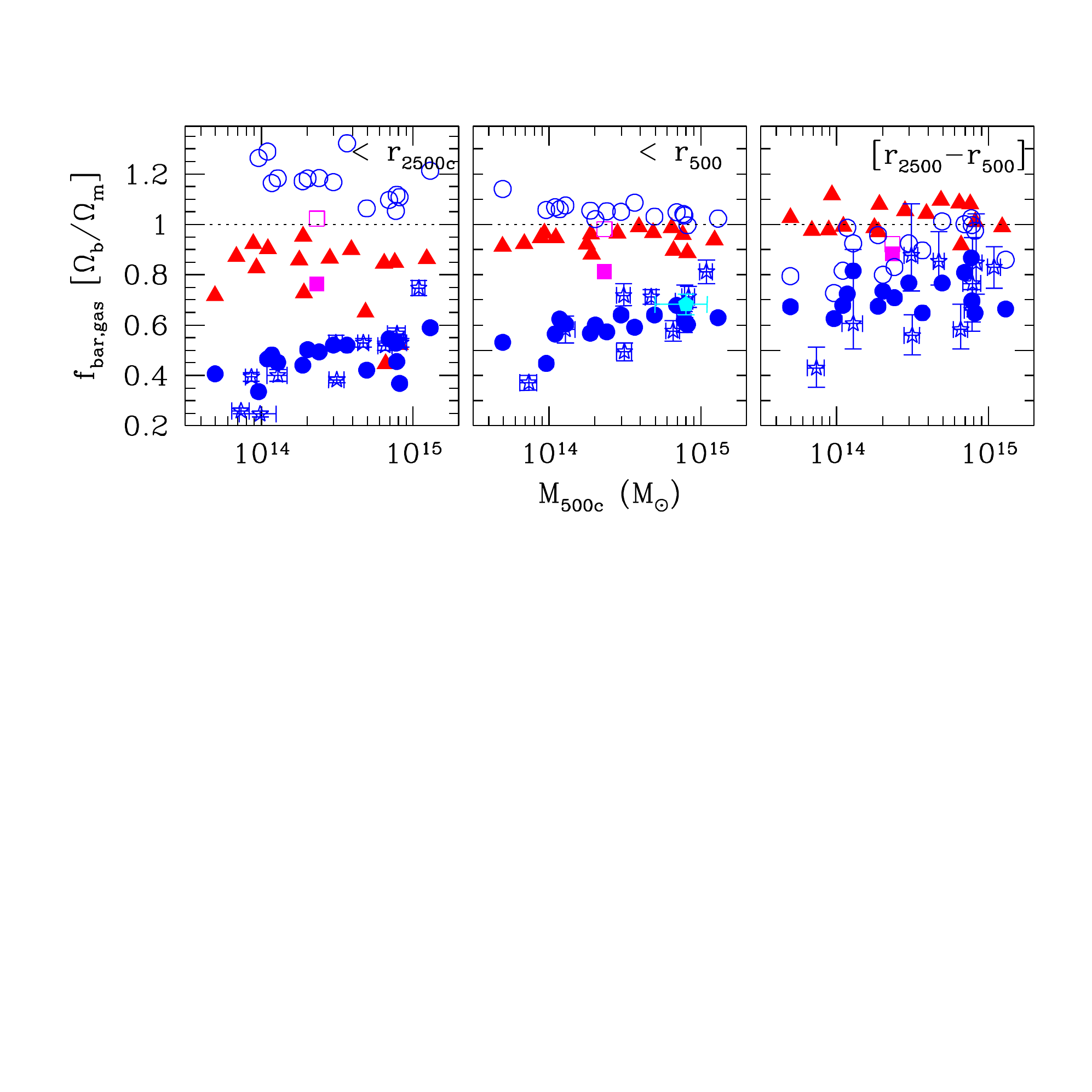}}
\vspace{-9cm}
\caption{The gas and total baryon fractions of individual clusters in
  units of the universal mean value within different radii (from the
  left to the right panels: $r<r_{2500}$, in the annulus
  $r_{2500}-r_{500}$, $r<r_{500}$) as a function of the cluster mass,
  $M_{500}$. {\it Stars} with error bars show the gas fractions
  measured for a sample of relaxed clusters using {\sl Chandra}
  observations \protect\citep{vikhlinin_etal06}, while cyan pentagon
  shows the average value of gas fraction measured in the XMM-DXL
  sample \protect\citep{2008A&A...482..451Z}. {\it Solid triangles}
  show the gas fractions in the adiabatic simulations. The {\it solid
    circles} show the gas fraction, while {\it open circles} show the
  total baryon fraction in the simulations with cooling and star
  formation. {\it Solid} and {\it open purple squares} show the gas
  and baryon fractions in the resimulation of one of the clusters, in
  which cooling was turned off at $z<2$. We use the universal baryon
  fraction of $\Omega_{\rm b}/\Omega_{\rm m}=0.1428$, assumed in
  cosmological simulations, to normalize the measured cluster gas and
  baryon fractions, and assume the universal value of $\Omega_{\rm
    b}/\Omega_{\rm m}=0.17$ for the observed clusters.}
\label{fig:fbar}
\end{figure}

Furthermore, the overall fraction of mass in the hot ICM and in the
stellar component can be used as a useful diagnostic of the heating
and cooling processes discussed in the previous section. Given the
depth of gravitational potential wells of clusters, the total mass in
baryons within a suitably large radius is expected to be close to the
mean mass fraction of baryonic matter in the Universe. Indeed,
non-radiative cosmological simulations show that the baryon fraction
is close to universal within the virial radius, and even at smaller
radii down to about half of the virial radius \citep[$\approx
r_{500}$; ][see also
Figure~\ref{fig:fbar}]{frenk_etal99,kravtsov_etal05,ettori_etal06}. The
actual value of the total baryon fraction, of the stellar and hot gas
fractions individually, and their scalings with cluster mass can be
used as indicators of the heating and cooling processes operating
during cluster formation.

Figure~\ref{fig:fbar} shows the total baryon (stars plus gas) and hot
gas mass fractions of observed and simulated clusters \citep[from the
sample presented in][]{nagai_etal07b} in units of the universal mean
value within different radii ($r_{2500}$ and $r_{500}$ correspond to
about a quarter and half of the virial radius, respectively). The red
triangles show the total baryon fraction in non-radiative
simulations. There is no visible trend of the baryon fraction with
cluster mass in these simulations \citep[see
also][]{crain_etal07}. The actual values are slightly below the
universal fraction because gas and dark matter exchange energy and
angular momentum during cluster formation with gas assuming a somewhat
more extended radial distribution than dark matter \citep{lin_etal06}.

In contrast, the baryon fraction in simulations with cooling and star
formation is larger than the universal fraction and there is some
trend of $f_{\rm bar}$ to decrease with increasing cluster mass.  The
enhanced baryon fraction is due to the fact that when baryons cool and
condense in the centers of halos, that merge to form a cluster, they
become much more resistant to tidal stripping than the dark matter,
which is relatively loosely bound and is spread out in extended halos.
Thus, compared to the non-radiative case, the condensed baryons are
able to reach the center of the cluster when massive halos merge while
dark matter is stripped and is deposited at larger radii. This
difference leads to an increase of the baryon fraction in the inner
cluster regions. In fact, the right panel clearly shows that the baryon
fraction in the annulus outside the core is not enhanced in runs with
cooling, but is actually somewhat suppressed.

Almost all of the baryons in the simulations shown in
Figure~\ref{fig:fbar} are either in the hot X-ray emitting gas or in
collisionless stellar particles, created out of cold condensed
gas. Thus, the difference between the total baryon fraction (open
circles) and the hot gas fraction (solid circles) in each panel gives
the stellar mass fraction. It is clear that the stellar fractions
within $r_{500}$ in these simulations are large (35--60 per cent depending
on halo mass and radius) compared to observational estimates of the
stellar mass fractions in clusters \citep[$f_{\ast}\approx
10-20$ per cent,][]{lin_etal03,gonzalez_etal07}.  This discrepancy is the
consequence of the overcooling problem discussed above (see
\S~\ref{sec:ch}).

Note that the overcooling problem and the cool core problem in
observed clusters, although perhaps related, are not necessarily the
same \citep{bryan_voit05}. The cool core problem is the {\it current}
lack of cooling gas in the cores of clusters where cooling time should
be short.  This requires balancing the radiative losses by suitably
heating the gas after the cluster potential well has formed, i.e. at
relatively low redshift, $z\mincir 1$.  The overcooling problem in
simulations is the problem of too high mass of cooled, condensed gas
produced {\it over the entire formation history} of the cluster,
including all of its progenitors. This problem is more serious as it
requires heating throughout the evolution of the cluster and of its
progenitors and in particular at high redshift ($z>2-3$), where most
of the cluster stars are born.

An additional puzzle is that the observed hot gas fractions
\citep[][see data points in
Figure~\ref{fig:fbar}]{vikhlinin_etal06,zhang_etal06,laroque_etal06}
are quite low and are in fact consistent with the hot gas fractions in
simulations with cooling. Taken at face value, the observational
measurements of stellar and gas fractions imply that the baryon
fraction in clusters is considerably smaller than the expected
universal value \citep[e.g.][]{ettori03,mccarthy_etal07b}. The open
and solid purple squares in the figure show baryon and hot gas
fractions in a re-simulation of one of the clusters, in which cooling
was artificially switched off after $z=2$ (i.e., in the 10 billion
years of evolution).  The stellar fraction in this simulation (the
difference between open and solid points) is about a factor of two
smaller than in the other runs with cooling. However, the hot gas
fraction is correspondingly larger and is larger than the observed gas
fractions. This means that suppressing the stellar fraction without
heating the gas, so as to push a significant fraction of it outside
$r_{500}$, is not a viable solution.

These considerations leave us with three possible options: 1) the
stellar fraction is suppressed by a heating mechanism, which also
causes a substantial decrease of the gas fraction (and, hence, of the
total baryon fraction) considerably below the universal value; 2)
observational estimates of stellar fraction are biased low by a factor
of 2-3, or a significant fraction of baryons ($\sim 20$--30 per cent)
in clusters, either in the form of a diffuse stellar component
\citep[e.g.,][]{murante_etal07,gonzalez_etal07} or of warm gaseous
phase, are missed in observations; 3) the hot gas fraction in
simulations are actually larger, and the low values inferred from
observations are biased low for some reason.

The first option is plausible given the expected heating by
supernova-driven winds and AGNs. The challenge in this case is to
explain how this heating can substantially reduce the gas fraction at
large radii in massive clusters (where the amounts of required heat
injection would be tremendous), while retaining a nearly self-similar
scaling of entropy at large radii (see Figure~\ref{kTx}). In
addition, explaining the very weak dependence of the gas fraction on
the total cluster mass within $r_{500}$ and in the shell between
$r_{2500}$ and $r_{500}$ \citep[see Figure~\ref{fig:fbar} and
][]{sun_etal08} is likely to be a challenge.

A detailed discussion of the second option is outside the scope of
this review. We simply note that at the present time there is no solid
observational evidence for a large fraction of baryons missing in
clusters.

The low bias in the measurements of hot gas appears to be unlikely if
the collisionless effects in the ICM plasma are negligible, as
detailed tests of observational analyses techniques used to estimate
gas fractions in the modern X-ray data shows that the methods are
quite accurate \citep{nagai_etal07a}. In fact, if there are
non-thermal sources of pressure support in clusters, such as cosmic
rays or turbulent motions, observational estimates of hot gas
fractions would be biased high (because hydrostatic mass estimates
using the thermal pressure along would be biased low).  However,
collisionless effects (see \S~\ref{sec:addphysics}) and deviations
from equilibrium in plasma, such as evaporation, can potentially lower
the gas fractions or bias observational estimates
\citep[e.g.,][]{loeb07}.

\subsection{Constraints on the models from the observed galaxy population}
Whatever the nature of the feedback mechanism that shapes the
thermodynamical properties of the ICM, we expect that it
leaves its imprint on the history of star formation and, therefore, on the
optical properties of the cluster galaxy population. In turn, a
crucial diagnostic to trace the past history of star formation within
cluster galaxies is represented by the pattern of chemical enrichment
of the ICM
\citep[e.g.,][]{1997ApJ...488...35R}. It has been 
known for over twenty years that the level of enrichment in
heavy elements of the ICM is about one third of the solar value, thus
demonstrating that a significant fraction of the intra-cluster gas has
been processed by stars and then expelled from galaxies
\citep[e.g.,][]{2005ApJ...620..680B}. Thanks to the much improved
sensitivity of the X--ray telescopes of the last generation, it is
also well established that the ICM metallicity, i.e. the ratio between
the mass in metals and the mass in hydrogen, is not uniform. Rather it
is enhanced in correspondence of the cool cores
\citep{2004A&A...419....7D,2005ApJ...628..655V,2007ApJ...666..835B,2008A&A...478..615S,2008SSRv..134..337W,2008A&A...487..461L},
where it reaches values comparable to solar. This enhancement is
generally interpreted as due to the contribution to the enrichment
from the past history of star formation in the BCG, thus
highlighting the interplay between galaxy evolution and properties of
the ICM.

\begin{figure}
  \centerline{ 
\hbox{
\psfig{file=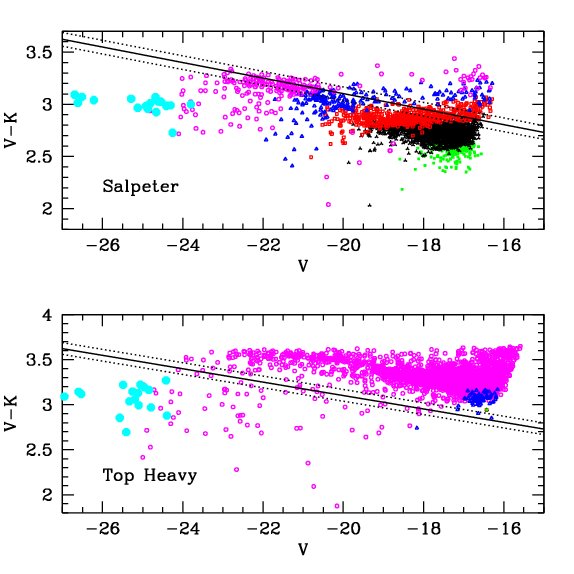,width=8.5truecm} 
\psfig{file=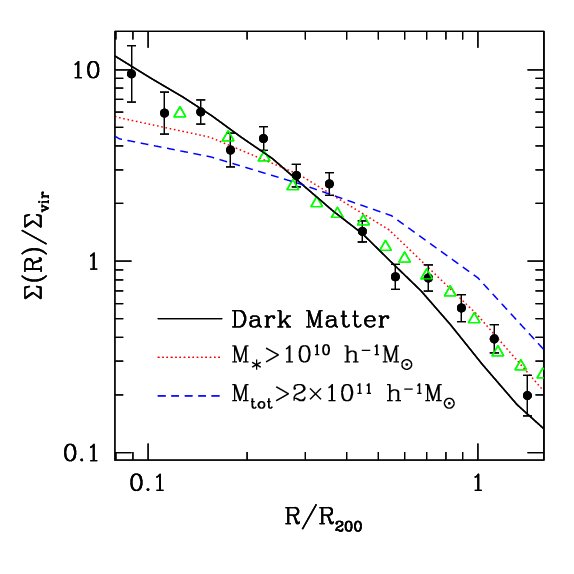,width=8.5truecm} 
}
}
\caption{Right panel: the $V$--$K$ vs. $V$ color--magnitude relation
  for a set of galaxy clusters simulated with the GADGET-2 code and
  including a description of chemical enrichment. The top and
  the bottom panels correspond to assuming a
  \protect\cite{1955ApJ...121..161S} IMF and a top--heavy IMF
  \protect\citep{1987A&A...173...23A}, respectively.  Straight lines in
  each panel show the observed CMR relations
  \protect\cite{1992MNRAS.254..601B}, with the corresponding intrinsic
  standard deviations. Big filled dots mark the BCG of each
  cluster. Different symbols and colors are used for galaxies having
  different metallicities. Magenta open circles: $Z>1.5Z_\odot$; blue
  filled triangles: $1.5<Z/Z_\odot<1$; red open squares:
  $1<Z/Z_\odot<0.7$; black open triangles: $0.7<Z/Z_\odot<0.4$; green
  filled squares: $Z<0.4Z_\odot$. Figure adopted from
  \protect\cite{2006MNRAS.373..397S} with copyright permission from
  2006 Blackwell Publishing Ltd. Left panel: projected radial
  distribution of galactic subhalos $\Sigma(R)/\Sigma_{vir}$ averaged
  over the eight simulated clusters at $z=0$ and three orthogonal
  projections.  We show the radial distribution of subhalos selected
  using the mass thresholds of $M_{\ast}>10^{10}h^{-1}M_{\odot}$ ({\it
    dotted}) and $M_{\rm tot}>2\times10^{11}h^{-1}M_{\odot}$ ({\it
    dashed}).  The solid line is the average projected profile of dark
  matter in the gasdynamics simulation.  The {\it solid circles} are
  the average radial profile of galaxies in clusters measured in two
  different surveys
  \protect\citep{1997ApJ...478..462C,2004ApJ...610..745L}.  The data
  points are scaled arbitrarily.  Note that the distribution of
  $M_{\ast}$-selected subhalos is consistent with the observed
  distribution of galaxies over the entire range of probed radii:
  $0.1<R/R_{\rm 200}<2.0$. Figure adopted from
  \protect\cite{nagai_kravtsov05} with copyright permission from 2005
  IOP Publishing Ltd.}
\label{fig:cmr}
\end{figure}

How can we properly model the process of metal enrichment and connect
it to star formation? Models of stellar evolution predict that SNe of
different types (i.e. SN-II and SN-Ia) arise from stars of different
mass, which explode over different time scales, and synthesize
different metal species in different proportions
\citep{2003ceg..book.....M}. Therefore, including a model of chemical
enrichment in simulations requires specifying the shape of the initial
mass function (IMF) for star formation, the lifetimes of stars of
different mass and the metal yields from different SN types \citep[see
][ for a review]{2008SSRv..134..379B}. Chemo--dynamical simulations of
galaxy clusters have shown that indeed metals are produced at a level
comparable with observations
\citep{valdarnini03,2006MNRAS.371..548R,2007MNRAS.382.1050T,2007A&A...466..813K,dave_etal08}. Also
the resulting spatial distribution of metals is not too different from
the observed one, although simulations tend to predict 
metallicity profiles in the central regions steeper than in
observations. This result is generally
interpreted as due to the excess of recent star formation taking place
in the core regions of simulated clusters \citep{2008MNRAS.386.1265F}.

Since galaxy colors are sensitive to both the star formation rate and
the metallicity of the stars, introducing an accurate description of
the chemical enrichment opens the possibility to reliably compare the
predicted and the observed properties of the cluster galaxy
populations. As we have already discussed, hydrodynamical simulations
treat the process of star formation through the conversion of cold and
dense gas particles into collisionless star particles, treated as
single age stellar populations (see \S~\ref{sec:techn}).  Luminosities
in different bands for each particle can be computed using suitable
stellar population synthesis models
\citep{2003MNRAS.344.1000B,1998ApJ...509..103S}. Galaxies in
simulations can then be identified as gravitationally-bound groups of
star particles by a group-finding algorithm
\citep[e.g.][]{1999ApJ...516..530K,2001PhDT........21S,SP01.2} and
their luminosity and colors calculated by summing up luminosities of
individual star particles.

These analyses provided results which are quite encouraging as for the
level of agreement with observations
\citep{2005MNRAS.361..983R,2006MNRAS.373..397S,2008MNRAS.389...13R}.
As an example, we show in Figure \ref{fig:cmr} the comparison between
the observed and simulated V-K vs K color--magnitude relation. The two
panels show the results of simulations for two sets of clusters, based
on assuming either a standard Salpeter IMF
(\citealt{1955ApJ...121..161S}, top panel) or a top--heavy IMF
(\citealt{1987A&A...173...23A}, lower panel) which predicts a much
larger number of massive stars and, therefore, of type-II SN. Quite
apparently, increasing the number of type-II SN increases the amount
of metals produced, with the result that the galaxy population becomes
too red. On the other hand, a standard IMF produces fairly realistic
galaxies, with a color sequence that reflects a metallicity sequence
(i.e. more metal rich galaxies have redder colors). While this is true
for the bulk of the galaxy population, simulated BCGs (indicated by
the large dots in Figure \ref{fig:cmr}) are much bluer than
observed. Again, this is the consequence of the excess of recent star
formation in central cluster regions and, as such, represents the
evidence in the optical band of the cool-core problem revealed by
observations in the X--ray band.

Although there are problems with reproducing luminosities and colors
in simulations, the overall spatial distribution of galaxies is
reproduced remarkably well
\citep[e.g.,][]{nagai_kravtsov05,maccio_etal06}, as is shown in the
right plot in Figure~\ref{fig:cmr}.  The fact that {\it ab initio}
cosmological simulations with simple phenomenological prescriptions
for converting gas into stars can match the observed radial
distribution of galaxies in clusters is a testament that the
hierarchical $\Lambda$CDM model of structure formation captures the
main processes operating during the formation of clusters.

\section{Summary and outlook for the future}
\label{sec:summary}

In the preceding sections we have discussed the techniques which are
currently used to carry out cosmological simulations of cluster
formation and reviewed recent progress in the numerical modeling of
galaxy clusters. Many of the salient observed properties of clusters,
such as scaling relations between observables and total mass, radial
profiles of entropy and density of the intracluster gas and radial
distribution of galaxies are reproduced quite well. In particular, the
outer regions of clusters at radii beyond about 10 per cent of the
virial radius are quite regular and exhibit scaling with mass
remarkably close to the scaling expected in the simple self-similar
model. However, simulations generally do not reproduce the observed
``cool core'' structure of clusters: simulated clusters generally
exhibit a significant amount of cooling in their central regions,
which causes both an overestimate of the star formation in the central
cluster galaxies and incorrect temperature and entropy profiles. The
total baryon fraction in clusters is below the mean universal value
and there are interesting tensions between observed stellar and gas
fractions in clusters and predictions of simulations. These puzzles
point towards an important role played by additional physical processes,
beyond those already included in the simulations: plasma transport
processes such as viscosity and conduction, AGN powering jets,
subsequent injection of turbulence and relativistic particles in
bubbles, whose evolution
depends on small-scale turbulence, magnetic fields and viscosity.

The observational studies of clusters are increasingly multi-wavelength
(IR, optical, X-ray, sub-mm/SZE, radio, gamma-rays). Cosmological
simulations of clusters will need to be prepared for the challenge of
modeling and interpreting the rich observational datasets which are
quickly becoming available. The inclusion of additional physical
processes required to reproduce the entire wealth of the data will
likely be driving the field of cluster simulations in the near
future. Although most of these processes and how they operate in
clusters are not well understood, there is a hope that they can be
constrained by observations themselves. In this sense, the
increasingly sophisticated numerical models of clusters together with
multi-wavelength observations serve as an astrophysical laboratory for
studying such processes.

As an example, the high-resolution X-ray spectroscopy and (possibly)
kinetic SZE will allow us to study the ICM velocity field and
constrain the amount and dynamical extension of turbulence.  Given
that simulations predict that turbulence is ubiquitous in clusters,
but its properties are sensitive to viscosity, such observations can
constrain viscous energy transport in the intracluster plasma. Large
collecting area X-ray telescopes of the next generation along with SZE
cluster surveys will allow us to trace evolution of the ICM thermal
and chemical properties out to high redshifts ($z>1-1.5$). This will
fill the gap between studies of low- to intermediate-$z$ ICM and
high-$z$ IGM and trace its evolution with redshift, thereby painting a
unified picture of diffuse baryons across a wide range of redshifts.

It may also turn out that too many rather complex processes are
shaping the properties
of the ICM in cluster cores, and detailed
modeling of these regions is fraught with too many uncertainties and
problems.  Including phenomenological prescriptions for modeling such
processes inevitably leads to the loss of the predictive power of the
simulations.  The practitioners modeling cluster formation (and galaxy
formation in general) should figure out how to balance the
uncertainties of the included processes with the goal of accurately
{\it predicting} details of observational data. Needless to say, it may
not be always possible. This is not a cause for despair as such
situations are very common in computational science. In fact, in many
problems in physics and astrophysics no exact numerical prediction is
possible. Simulations, however, can be very useful even in this case
because they can illuminate the ways in which certain processes operate, 
thereby providing a powerful insight into
the physics of the problem and motivating and guiding development of
further models \citep[see, e.g.,][for insightful discussion of this
issue]{kadanoff04}
\footnote{Available at 
${\tt
    http://jfi.uchicago.edu/\sim leop/AboutPapers/Computational\_Scenarios.pdf}$
}.

\vskip 2truecm

{\bf{\Large{Acknowledgments}}}

\bigskip

We are grateful to Nick Gnedin, Giuseppe Murante and Romain Teyssier
for useful discussion of the plasma effects in the intracluster
medium, to Gianfranco Brunetti and Klaus Dolag for discussion on the
role of cosmic rays, to Daisuke Nagai for providing Figure~\ref{kTx}
and   comments on the draft of this manuscript and to Dan
Marrone for providing SZA image of Abell 1914. SB acknowledges
financial support of the PRIN-MIUR grant under the project ``The
Cosmic Cycle of Baryons'', of a ASI-AAE Theory Grant and of the
PD51-INFN grant. AK is supported by the NSF grants AST-0507666, and
AST- 0708154, by NASA grant NAG5-13274, and by Kavli Institute for
Cosmological Physics at the University of Chicago through grant NSF
PHY-0551142 and an endowment from the Kavli Foundation. AK wish to
thank the Institute for Theoretical Physics at University of Z\"urich
for hospitality during the completion of this review. This work made
extensive use of the NASA Astrophysics Data System and arXiv.org
preprint server.

\bibliography{review,bibliography,review_sb}

\end{document}